\documentclass[usenatbib]{mn2e}
\usepackage{graphicx}
\usepackage[usenames]{color}
\usepackage{times}
\def\gtsim{\lower.5ex\hbox{$\; \buildrel > \over \sim \;$}}

\title[UV Upturn Galaxies]{Galaxy And Mass Assembly (GAMA): Defining Passive Galaxy Samples and Searching for the UV Upturn}
\author[S. Phillipps et al.]
{S. Phillipps$^1$, S.S. Ali$^{1,2}$, M.N. Bremer$^1$, R. De Propris$^3$, A.E. Sansom$^4$, M.E. Cluver$^{5,6}$, 
\and M. Alpaslan$^{7}$, S. Brough$^{8}$, M.J.I. Brown$^9$, L.J.M. Davies$^{10}$, S.P. Driver$^{10,11}$, 
M.W. Grootes$^{12}$, 
\and B.W. Holwerda$^{13}$, A.M. Hopkins$^{14}$, 
P.A. James$^{15}$, K. Pimbblet$^{16}$, A.S.G. Robotham$^{10}$,
\and  E.N. Taylor$^5$ and L. Wang$^{17,18}$
\\
$^1$Astrophysics Group, School of Physics, University of Bristol, Tyndall Avenue, Bristol, BS8 1TL, UK\\
$^2$Subaru Telescope, NAOJ, 650 North A'ohoku Place, Hilo, HI 96720, USA\\
$^3$ Finnish Centre for Astronomy with ESO, University of Turku, Vesilinnantie 5, 21400, Turku, Finland\\
$^4$Jeremiah Horrocks Institute, School of Physical Sciences and Computing, University of Central Lancashire, Preston PR1 2HE, UK\\
$^5$Centre for Astrophysics and Supercomputing, Swinburne University of Technology, Hawthorn, VIC 3122, Australia\\
$^6$Department of Physics and Astronomy, University of the Western Cape, Robert Sobukwe Road, Belville 7535, South Africa\\
$^{7}$Center for Cosmology and Particle Physics, Department of Physics, New York University, NY 10012, USA\\
$^{8}$School of Physics, University of New South Wales, NSW 2052, Australia\\
$^9$School of Physics and Astronomy, Monash University, Clayton, Victoria 3800, Australia\\
$^{10}$ICRAR, University of Western Australia, 35 Stirling Highway, Crawley, WA 6009, Australia\\
$^{11}$SUPA, School of Physics and Astronomy, University of St Andrews, North Haugh, St Andrews, Fife KY16 9SS, UK\\
$^{12}$Netherlands eScience Center, Science Park 140, 1098 XG Amsterdam, The Netherlands\\
$^{13}$Department of Physics and Astronomy, 102 Natural Science Building, University of Louisville, Louisville KY 40292, USA\\
$^{14}$Australian Astronomical Optics, Macquarie University, 105 Delhi Road, North Ryde, NSW 2113, Australia\\
$^{15}$Astrophysics Research Institute, Liverpool John Moores University, IC2, Liverpool Science Park, 146 Brownlow Hill, Liverpool L3 5RF, UK\\
$^{16}$E.A. Milne Centre for Astrophysics, University of Hull, Cottingham Road, Kingston-upon-Hull HU6 7RX, UK\\
$^{17}$SRON Netherlands Institute for Space Research, Landleven 12, 9747 AD Groningen, The Netherlands\\
$^{18}$Kapteyn Astronomical Institute, University of Groningen, Postbus 800, 9700 AV Groningen, The Netherlands
}
\begin{document}

\date{Accepted . Received ; in original form }

\pagerange{\pageref{firstpage}--\pageref{lastpage}} \pubyear{}

\maketitle

\label{firstpage}

\begin{abstract}
We use data from the GAMA and GALEX surveys to demonstrate that the UV upturn, an unexpected excess of ultraviolet flux from a hot stellar component, seen in the spectra of many early-type galaxies, arises from processes internal to individual galaxies with no measurable influence from the galaxies' larger environment. We first define a clean sample of passive galaxies without a significant contribution to their UV flux from low-level star formation. We confirm that galaxies with the optical colours of red sequence galaxies often have signs of residual star formation, which, without other information, would prevent a convincing demonstration of the presence of UV upturns. However, by including (NUV$-u$) and {\it WISE} (W2-W3) colours, and FUV data where it exists, we can convincingly constrain samples to be composed of non-star-forming objects. Using such a sample, we examine GALEX photometry of low redshift GAMA galaxies in a range of low-density environments, from groups to the general field, searching for UV upturns. We find a wide range of (NUV$-r$) colours, entirely consistent with the range seen -- and attributed to the UV upturn -- in low-redshift red sequence cluster galaxies. The range of colours is independent of group multiplicity or velocity dispersion, with isolated passive galaxies just as likely to have blue UV-to-optical colours, implying significant upturn components, as those in richer groups and in the previous data on clusters. This is supported by equivalent results for (FUV$-r$) colours which are clear indicators of upturn components. 

\end{abstract}


\section{Introduction}
The `UV upturn' signifies the {\it a priori} unexpected excess of ultraviolet flux seen in the spectra of many early-type galaxies, compared to what would be expected for a conventional old, metal-rich stellar population \citep{CodeWelch1979,Bertola1982,Oconnell1999}. The upturn is therefore commonly attributed to a minority old but hot stellar population, such as hot horizontal branch stars \citep{Greggio1990,Yi1998}, with a helium-enhanced population often being posited \citep[e.g.][]{Norris2004,Lee2005,Chung2011}. Although \cite{Burstein1988} early on reported an upturn in several nearby fairly isolated ellipticals such as NGC~4697, almost all subsequent work concentrated on early-type galaxies in rich clusters such as Virgo \citep{Boselli2005} or Coma \citep{Smith2012}, and generally on only the brighter cluster members \citep[e.g.][]{Brown2000,Brown2003}.

More recently \cite{Ali2018a,Ali2018b,Ali2018c} have demonstrated that the upturn is common in early-type galaxies in clusters across a broad range of galaxy luminosities, in a wide variety of clusters with different physical properties, at different redshifts \citep[see also][]{Ali2019}. This appears to indicate that the phenomenon is internal to the stellar populations of many individual passive galaxies, with a range of (reasonably old) ages, with no obvious environmental effects at play. In their work, they created reasonably well-sampled ultraviolet spectral energy distributions (SEDs) from a combination of photometry from different sources (GALEX, UVOT and SDSS). The spectral shapes which they determined then enabled them to rule out (significant) contributions to the UV flux from star formation in their sample of objects and hence demonstrate the presence of upturn components. However, for most large-scale galaxy surveys, this level of detail in the UV SED will not generally be available. We will therefore need to eliminate star-forming galaxies from samples of potential (passive) upturn galaxies by other means, such as using multiple broad-band colours.
 
In the present paper we use low redshift ($z < 0.05$) galaxies from the Galaxy And Mass Assembly (GAMA) survey \citep{Driver2011,Liske2015,Baldry2018} to explore the best means of extracting purely passive galaxy samples via the wide range of available broad-band photometry. We then use such a sample to investigate whether UV upturns exist in the spectra of early type galaxies in groups and the field and, if so, whether they have the same range of strengths, as represented by near- and far-UV to optical colours, as early types in clusters. NUV and FUV data for our sample galaxies are obtained from the GALEX catalogues, which have been previously cross-matched to the GAMA catalogues as described in \cite{Liske2015}.

We further take advantage of the rich panchromatic survey data in GAMA \citep{Driver2016} to explore the characteristics of (optical) red sequence galaxies in general, and those which are candidates for totally passive early-types with UV upturns in particular.  Besides the observed and rest-frame magnitudes across a wide range of bands, including particularly the mid-infrared from {\it WISE} \citep{Cluver2014} and far-infrared from {\it Herschel} \citep{Eales2015}, the GAMA database provides fits to the multi-wavelength SEDs made using stellar population synthesis techniques \citep{Taylor2011,Driver2016,Wright2016}.

All magnitudes used in this work are in the AB system. Where relevant we use H$_0$ = 70~kms$^{-1}$Mpc$^{-1}$, $\Omega_{\rm m} = 0.3$ and $\Omega_\Lambda = 0.7$ as in \cite{Taylor2011} from whose (updated) catalogue we take our basic GAMA parameters.

\section{Sample selection}
\label{samp}
The GAMA survey is based on a highly complete galaxy redshift survey \citep{Baldry2010, Hopkins2013, Liske2015, Baldry2018} covering approximately 280 deg$^2$ to a main survey magnitude limit of $r < 19.8$. This area is split into three equatorial (G09, G12 and G15) and two southern (G02 and G23) regions. In the present work we use galaxies from the three equatorial fields. The spectroscopic survey was undertaken with the AAOmega fibre-fed spectrograph \citep{Saunders2004, Sharp2006} allied to the Two-degree Field (2dF) fibre positioner on the Anglo-Australian Telescope \citep{Lewis2002}. It obtained redshifts for $\sim 300,000$ targets covering $0<z <0.6$, with a median redshift of $z\simeq 0.2$, with high (and uniform) spatial completeness (98.5\%) on the sky in the GAMA equatorial areas \citep{Robotham2010}. A total of 195,669 galaxies have high quality spectra (GAMA redshift quality code nQ$>2$) in the GAMA catalogues

Built around the redshift survey, photometric data are provided at a wide range of wavelengths from the far-ultraviolet to far-infrared. Full details can be found in \cite{Driver2011, Driver2016} and \cite{Liske2015}.

Within the multi-wavelength database, each galaxy is also characterised by a wide range of  derived parameters. Those of interest in the current work include rest-frame magnitudes and colours, stellar masses and other stellar population parameters, derived from multi-wavelength SED fitting \citep{Taylor2011}. The fits themselves are made across the {\it rest-frame} wavelength range 3000-10000 \AA. They assume \cite{Bruzual2003} models with a \cite{Chabrier2003} stellar initial mass function and with a variety of possible metallicities, and include allowance for the effects of dust extinction via a \cite{Calzetti2000} extinction curve. Further interstellar medium (ISM) and star formation properties are derived via a MAGPHYS \citep{daCunha2008} analysis of a wide range of photometric measures from the far-UV to far-IR \citep{Driver2016, Wright2016}.

For comparison with low redshift cluster data \citep{Ali2018a,Ali2019}, we select galaxies with \citep[flow-corrected:][]{Tonry2000,Baldry2012} redshifts $0.002 < z <0.05$. As we wish to study moderate-to-high luminosity objects (again for comparison with the local cluster samples) we choose to limit luminosities to $M_r < -19$. Note that this also ensures that we have a securely volume complete sample; GAMA is spectroscopically complete to below $M_r = -17.5$ at our maximum distance. Furthermore, the combination of low redshift and high luminosity provides the best prospect of the required ultraviolet and mid-infrared photometry having small errors. At these redshifts k-corrections in the UV  are small enough to be neglected relative to the photometric errors \citep[less than about 0.1 mags, see][]{Ali2019, Kaviraj2007a}, while k-corrections in the optical are already taken into account in the data.  Look-back times are all less than 0.7~Gyr, so evolutionary effects across the sample depth will be very small: at moderate to old ages the \cite{Conroy2009} models used by \cite{Ali2018a}, for instance, evolve by only 0.03 magnitudes in ($U-V$) per Gyr \citep[see also][]{Phillipps2019}.

To explore the UV upturn we require at least a 5-sigma detection of a given galaxy (equivalent to a magnitude error less than 0.2) in the GALEX NUV band at central wavelength 232~nm; the FUV band observations at 154~nm are less deep \citep{Morrissey2007}. The matching of GAMA and GALEX detections is discussed in detail in Section 4.2 of \cite{Liske2015}, who demonstrate that the GALEX detection rate (for typical exposure times of 1500~s) is better than 80\% for low-redshift galaxies of all types brighter than $r = 18$, which covers all our galaxies. With the 5-sigma limit, this leaves us with 797 GALEX objects.

As we are interested in the environments of our galaxies \citep{Brough2013}, we also match them to the GAMA group catalogue \citep[G3C:][]{Robotham2011}. This further requirement leaves us with a sample of 773 nearby bright NUV-detected galaxies  (after one object with an implausible ($g-r$) colour is removed). Structural parameters from S\'{e}rsic profile fits \citep{Kelvin2012} and morphological type classifications \citep{Kelvin2014,Moffett2016} are also utilised.

\section{The Red Sequence}

\begin{figure}
\includegraphics[width=\linewidth]{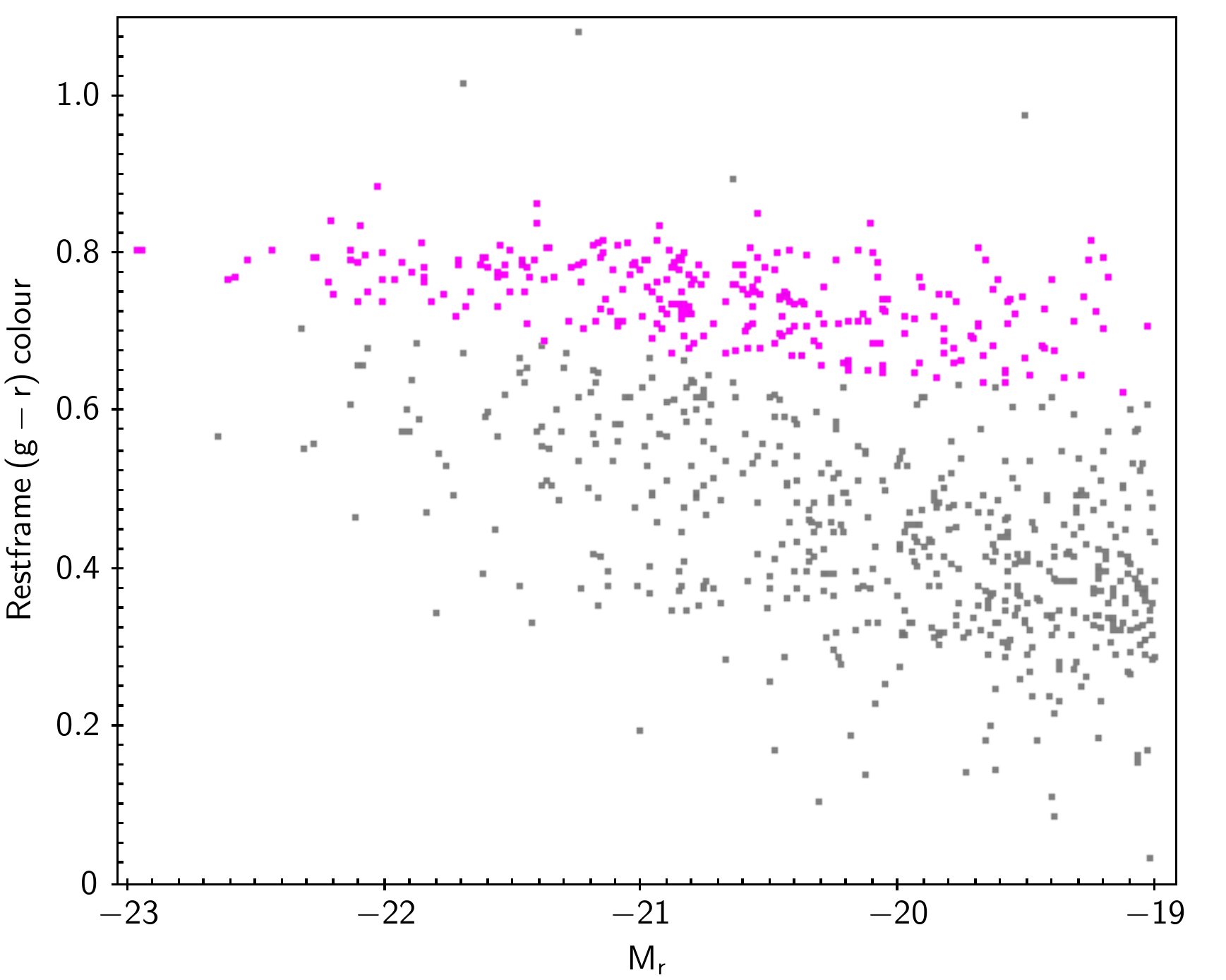}
\caption{Standard optical rest-frame ($g-r$) versus $M_r$ colour magnitude diagram for our overall sample of 773 nearby bright galaxies. The objects picked out in pink comprise the initial optical red sequence selection of 265 galaxies, grey points are the remainder, i.e. essentially blue cloud objects. Photometric errors in ($g-r$) are less than 0.05 magnitudes and those in $M_r$ are around 0.02 magnitudes.}
\label{gmr_r_cmd}
\end{figure}

The standard procedure for selecting passive galaxies is to use the split between the `red sequence' and `blue cloud' (e.g. Baldry et al. 2004 {\it et seq.}) in an optical colour-magnitude diagram, e.g. ($g-r$) versus $M_r$. We display this for our sample galaxies in Fig. \ref{gmr_r_cmd}. Note that $g$ and $r$ have been corrected to the rest frame but have not been corrected for any internal dust reddening. The slope of our red sequence ($\simeq -0.03$) is consistent with numerous other local determinations and the width of the region chosen is $\pm 0.1$ magnitudes, following \cite{DePropris2017} and \cite{DePropris2018}. This optical red sequence sample contains 265 galaxies. 

The UV upturn in passive galaxies is conventionally quantified by a UV-to-optical colour, usually (NUV$-r$) or (FUV$-r$) as in, e.g., \cite{Smith2012} and \cite{Ali2019}. Although using the FUV is preferable in some ways \citep[having a longer lever arm into the upturn region, e.g.][]{Dorman2003}, coverage is sparser and (NUV$-r$) is still a good proxy for the upturn strength. The upturn component contributes significant light, alongside the normal old stars, at least to 2800\AA $\;$ \citep{Burstein1988,Dorman1995,Smith2012} (recall that our NUV band is around 2300\AA). \cite{Burstein1988} originally discussed the steeply rising flux below 2000\AA$\:$ in some early type galaxies, but also noted that this hot component affected the whole spectrum shortward of 3200\AA. In general terms, the minimum in the SED around 2500\AA $\:$ in typical passive galaxy spectra does not indicate that the hot component is unimportant at these wavelengths, but that this is the cross-over point between the steeply declining old component and the steeply rising hot component, i.e. where the two components are roughly equally important \citep[see e.g.][figure 1]{Yi1998, Dorman2003}. A detailed consideration of the spectral components, as in \cite{Ali2018a}, using models of old, high metallicity populations from \cite{Conroy2009}, implies that there is relatively little flux from the old component below 2800\AA $\:$ \citep[see also][]{Dorman2003}. \cite{Dorman1995, Dorman2003} noted that for strong UV sources (with no star formation) the contribution of the hot (horizontal branch) population to the 2500\AA $\:$ flux is around 75-80\%.

\begin{figure}
\includegraphics[width=\linewidth]{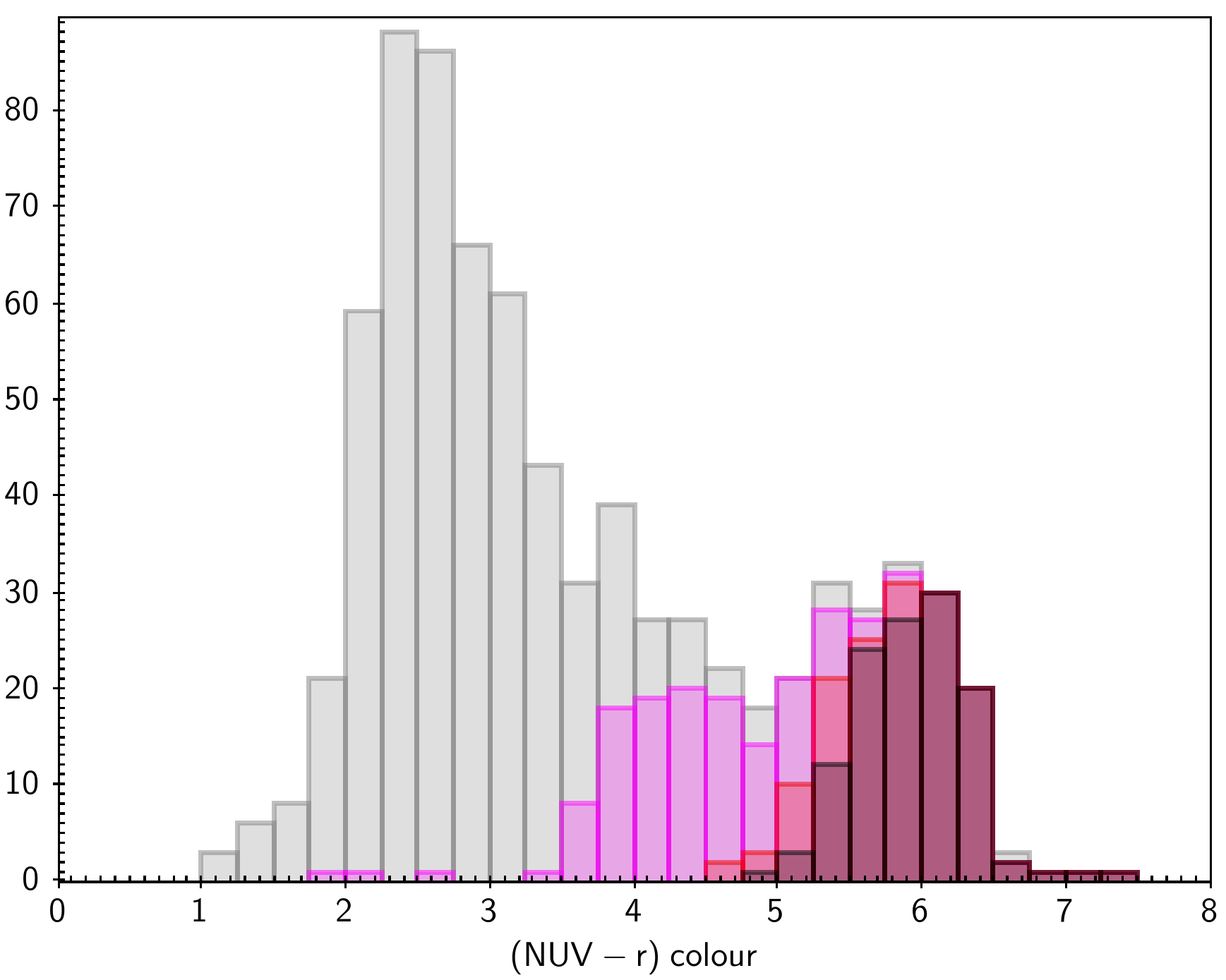}
\caption{Distribution of rest-frame (NUV$-r$) colours for our overall sample of nearby bright galaxies (grey) and the initial optical red sequence selected sample of galaxies (pink). The colour errors are around 0.2 magnitudes. The red histogram shows the distribution for the red sequence galaxies after the use of the additional cut in (NUV $-u$) as discussed in Section 3.1 (the NUV red sequence sample with 147 objects) and the dark red (black bordered) histogram shows the 122 objects remaining after a further cut on {\it WISE} (W2$-$W3) colour (the NUV+{\it WISE} red sequence sample from Section 3.2).}
\label{NUV_optsel_hist}
\end{figure}

Fig. \ref{NUV_optsel_hist} shows the distribution of (NUV$-r$) colour for our total sample (in grey) and for the optical red sequence galaxies (pink). It is immediately apparent that, as is well known, the optical red sequence is significantly contaminated by galaxies with the NUV-optical colours of star-forming galaxies, i.e. (NUV$-r$) less than about 5 \citep[see e.g.][]{Kaviraj2007a,Kaviraj2010,Crossett2017}. These are interpreted as early-type galaxies with residual star formation \citep[e.g.][]{SalimRich2010} or `red spirals' with strongly suppressed star formation \citep[e.g.][]{Crossett2014}.\footnote{Optical red sequence galaxies may also include interlopers with significant star formation but strong dust reddening \citep[e.g.][]{Sodre2013}, which will be red in all colours.}

This already indicates that, following \cite{Yi2005}, \cite{SalimRich2010} and \cite{Salim2014} for instance, we can obtain a cleaner passive sequence by cutting at around (NUV$-r$) = 5, where we see a clear minimum in the distribution of UV-optical colours. \cite{Crossett2014} use a stricter cut at (NUV$-r$) = 5.4 to separate out galaxies in rich clusters with residual star formation.

In the specific context of upturn galaxies, \cite{Yi2011} required a combination of (NUV $-r) > 5.4$, (FUV-NUV) $< 0.9$ and (FUV $-r) < 6.6$ in order for a galaxy to count as non-star forming with an upturn. However, this -- and specifically the very blue (FUV-NUV) limit -- appears to be based on a particular assumption for the temperature of the stars contributing to the UV upturn and returned no candidates in the Coma cluster, clearly in disagreement with other work. Based on the spectra of apparently passive galaxies, \cite{Smith2012} suggest that a `continuous' consideration of the colours is to be preferred, rather than Yi et al.'s imposition of discrete limits \citep[see also][]{Ali2018a}, with the (FUV-NUV) colours of their passive objects ranging up to about 2 \citep[see also][]{Brown2014}. \cite{Rich2005} had previously also found a wide range of (FUV-NUV) for (spectroscopically) quiescent early type galaxies \citep[see also][]{Donas2007,Carter2011}. On the other hand, Rich et al.'s wide range of (FUV$-r$) colours for quiescent galaxies strongly overlapped with those of star forming galaxies at (FUV$-r) < 6$ (their Figure 1), suggesting that residual effects of low-level star formation can be present even when no measurable emission lines are evident.

\cite{Arnouts2013} compared specific star formation rates (sSFRs) for their $z<0.2$ GALEX and SWIRE detected galaxies, as derived from UV to mid-IR SED fitting, to their positions in a (NUV$-r$) versus ($r-K$) colour-colour diagram. They found that galaxies with red (NUV$-r$) for their ($r-K$) had a low mean sSFR, with a clear subset of their galaxies falling in this quadrant, well separated from the blue star forming galaxies. However, while centred on (NUV$-r) \simeq 5$ their selection included objects as blue as (NUV$-r) \simeq 4$ and only removed objects with sSFR $< 10^{-10.5}$yr$^{-1}$, which is insufficiently tight if we wish to isolate the effects of a genuine UV upturn. \cite{Rawle2008} had similarly used NUV and the infrared $J$ band, again finding a wide colour spread, though not associating it with an upturn component. \citep[See also][for a recent extension of the UBJ method, though for high $z$ galaxies]{Leja2019}.

In any case, the existence of two roughly Gaussian distributions in the bimodal overall (NUV$-r$) colour distribution cautions \citep[cf.][]{Taylor2015} that some objects with residual star formation may still exist at (NUV$-r) > 5$. With these considerations in mind, in order to investigate further the selection of truly passive galaxies (with or without UV upturns), we next consider additional constraints from a broad range of other colours.

\subsection{The UV Spectral Slope}

First we could consider simply using a selection on ($u-r$), which is known to correlate reasonably well with sSFR \citep[e.g.][]{Bremer2018}. However, in practice this makes virtually no difference, the ($g-r)$ selected objects remain on the ($u-r$) red sequence \citep[cf. the tight correlation of $(u-g$) with ($g-r$) for early-type galaxies in Figure 5 of][for instance]{Brown2014}. We therefore need to consider shorter wavelengths than $u$, where the sensitivity to star formation is greater \citep[e.g.][]{Schawinski2007}.

In general terms, the SED of non-star-forming `upturn galaxies' will have red optical colours and an inflection below the $u-$band, before rising again (or, strictly, changing the SED slope upwards) at shorter wavelengths. By comparison, completely passive galaxies with no upturn population will have strongly declining flux from the $u$ to the NUV. Finally, any star-forming component will have, or add, a roughly flat spectrum at short wavelengths, while the galaxy's optical colours will be generated by the combination of young and old populations \citep[cf. the various model spectra in][for instance]{Hernandez2014}. Together, this implies that, for upturn galaxies, we should be looking for galaxies which are bluer in (NUV$-r$) or (NUV$-u$) than the standard passive models, yet not so blue as galaxies with (even small amounts of) star formation.

\begin{figure}
\includegraphics[width=\linewidth]{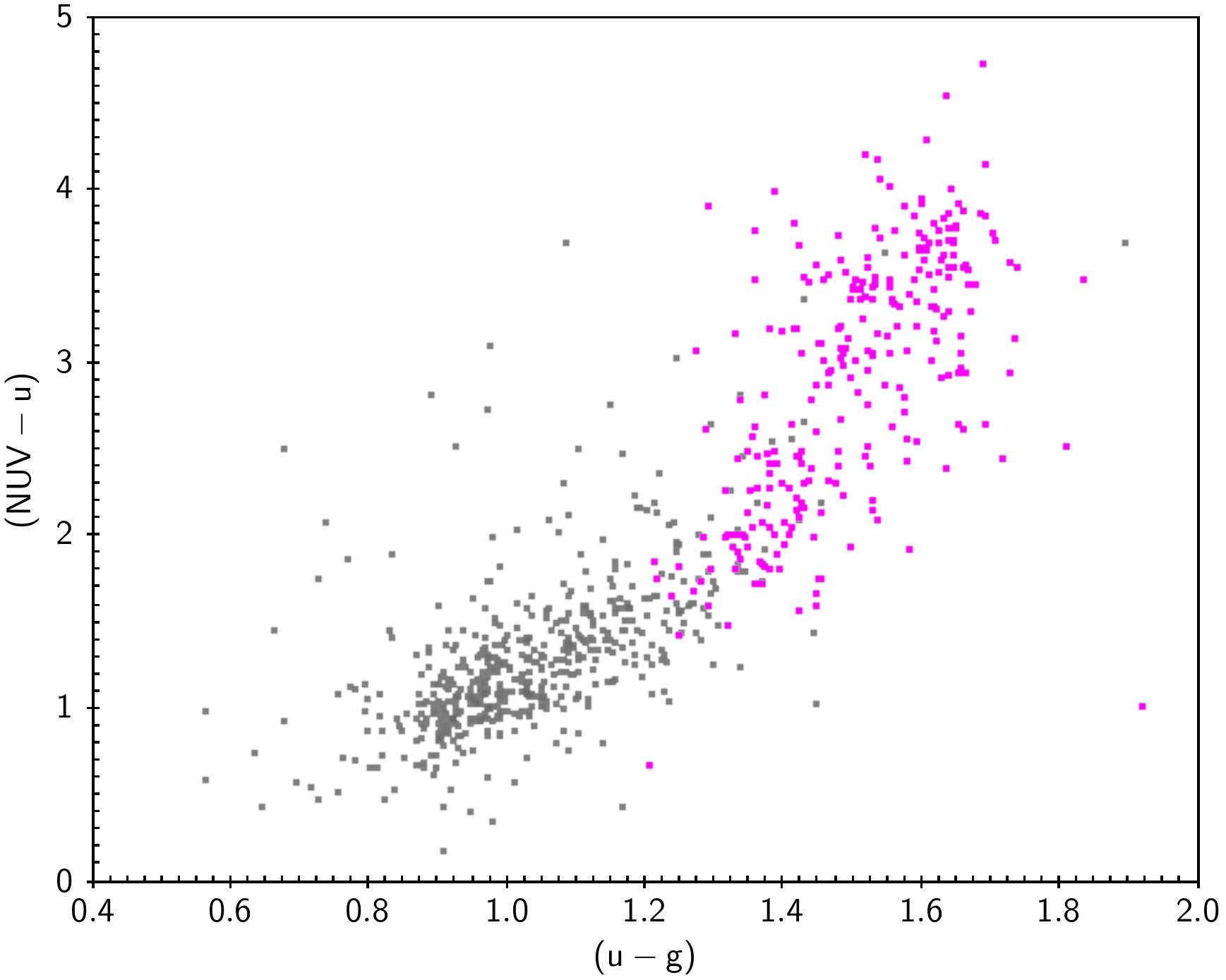}
\includegraphics[width=\linewidth]{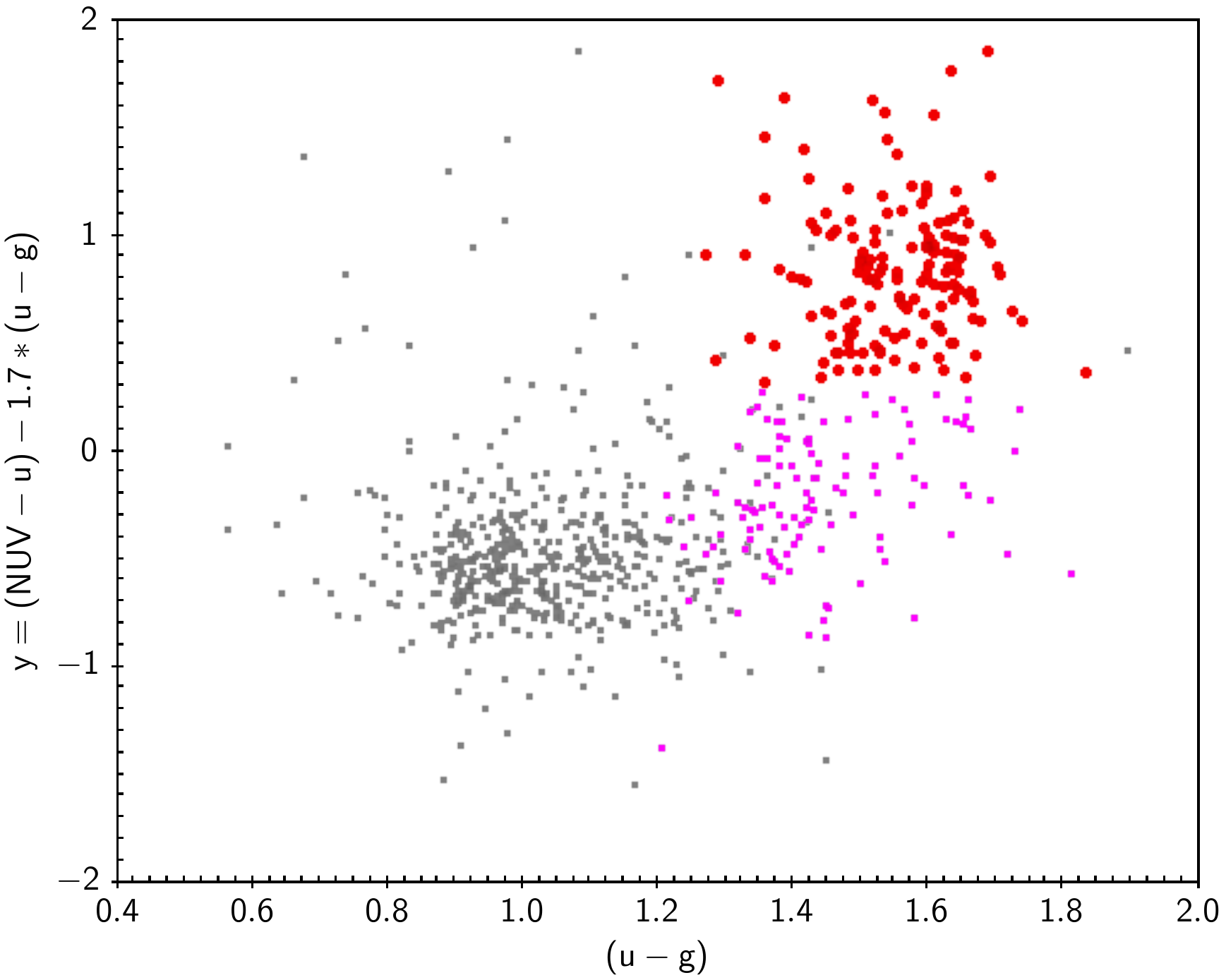}
\caption{Top panel: The distribution of rest-frame (NUV$-u$) versus ($u-g$) colours for our overall sample (grey). The objects picked out in pink comprise the initial optical red sequence selection. Bottom panel: Same but using the `rectified' colour $y$, defined as the (NUV$-u$) colour relative to its trend with ($u-g$) for star-forming galaxies (so that star-forming galaxies have the same mean $y$ at all ($u-g$), see text). Galaxies marked by red circles show our more stringent NUV red sequence selection. Photometric errors are around 0.07 magnitudes in ($u-g$) and 0.2 magnitudes in (NUV$-u$).}
\label{NUV_u}
\end{figure}

In Fig. \ref{NUV_u} we therefore plot, in the top panel, the diagnostic colour-colour diagram (NUV $-u$) versus ($u-g$) for our whole sample (grey), and for the optical red sequence (pink). This essentially compares the slope of the SED at wavelengths above and below the $u$-band. In the bottom panel, we show the `rectified' version, where we have removed the overall trend for the non-red sequence (i.e. star-forming) galaxies by plotting $y=$ (NUV$-u) - 1.7(u-g)$.

It is evident that `blue cloud' galaxies actually follow a quite well-defined sequence in this colour-colour plot \citep[cf.][]{Salim2007}, and that the extension of the sequence to redder ($u-g$) colours (which can reasonably be assumed to represent a lower, but non-zero, fraction of star formation) clearly runs into the area where optical red sequence, but `NUV blue', objects lie.

We can therefore use these plots to improve our selection of truly passive objects. As a first cut we could remove from the red sequence those galaxies below $y=0$ (the top of the main distribution for the `blue sequence'). However, as a more cautious/stringent cut, we choose to also remove galaxies slightly above the bulk of star formers (perhaps shifted there through photometric errors, which are typically 0.2 magnitudes) via a cut at $y=0.3$. Objects passing this cut are shown in red in the lower panel of Fig. \ref{NUV_u}. We will refer to the remaining subsample as the NUV red sequence galaxies. This subsample contains 147 of the original 265 optical red sequence galaxies. 

To demonstrate the effect of the additional (NUV$-u$) cut, Fig. \ref{NUV_optsel_hist} above, also shows (in red) the histogram of the (NUV$-r$) colour for the NUV-selected red sequence. It is clear that making the additional cut in (NUV$-u$) vs. ($u-g$) space has removed virtually all the objects with (NUV$-r$) below 5.0, which {\it post hoc} justifies the standard cut at the overall minimum in the (NUV$-r$) distribution when selecting genuinely passive galaxies \citep{Salim2014, Ali2019}.  Note particularly that (NUV$-r$) itself is {\em not} used in the selection at any point in this process (though of course (NUV$-u$) is). We can also note that the very few remaining galaxies with (NUV$-r) < 5$ in the NUV-selected red sequence are all at the faint end of our sample, with $M_r > -19.8$). This is likely due to the upturn component sitting on top of a lower metallicity old population (see Section 5, below). 


Although we expect this cut to do a reasonably good job of excluding galaxies with residual star formation, it is obvious that {\em arbitrarily small} amounts of star formation will cause very small shifts from the colours of totally passive galaxies, so can {\em never} be ruled out from a single colour, or indeed a single colour-colour plot. If we add small but gradually increasing amounts of a star forming (flat spectrum) component to an assumed completely passive galaxy at the top right of the distribution in Fig. 3, it will initially move essentially vertically downwards (increase in NUV flux but no measurable effect on $(u-g)$, and then diagonally down and to the left, following the overall distribution that we, indeed, see for the pink points, consistent with them representing galaxies with small but measureable star formation, as we have assumed (see also Section 3.3 below).
 
As further confirmation that our chosen cut is appropriate, we have examined the (NUV$-u$) colours of the Coma Cluster early type galaxies with well-determined upturns from the multi-band ultraviolet SED fitting of \cite{Ali2018a}. We find that the bluest colours among their upturn galaxies are (NUV$-u$) $\simeq 2.85$, corresponding to $y \simeq 0.3$, in excellent agreement with where we placed our cut in Fig. 3.

\subsection{Mid-infrared Colours}

\begin{figure}
\includegraphics[width=\linewidth]{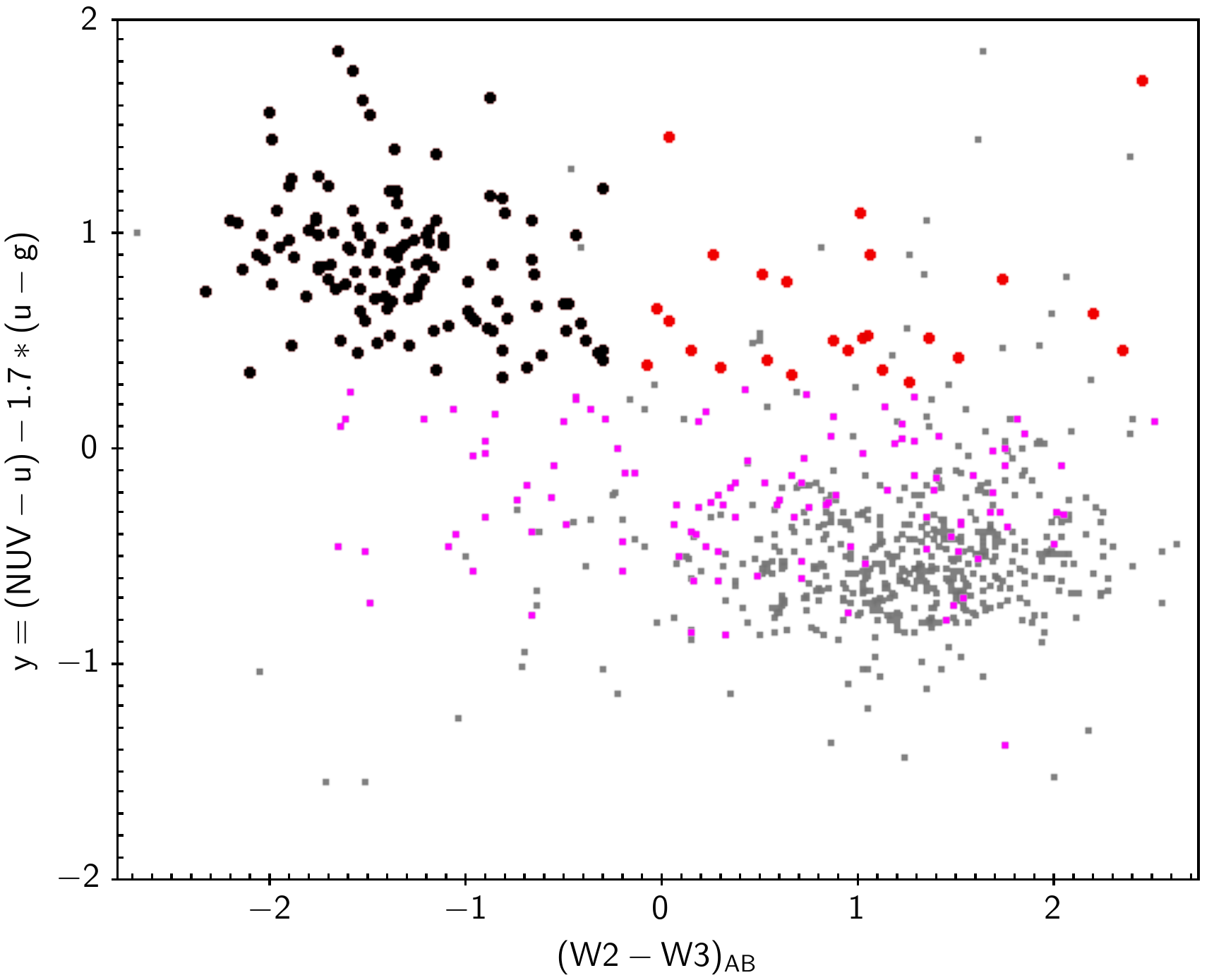}
\caption{Rectified NUV-optical colour $y$ versus {\it WISE} (W2-W3)$_{AB}$ colour for the various samples, plotted with the same colour codes as before. In addition the 122 objects from the NUV red sequence which also have {\it WISE} colours of non-star-forming galaxies are now plotted in black (the NUV+{\it WISE} red sequence sample). Although we do not impose a strict upper limit, errors in the {\it WISE} colours for our galaxies are less than 0.2 magnitudes in the large majority of cases. Errors in $y$ are also around 0.2.}
\label{NUVrs_wise23}
\end{figure}

We next check whether there may still remain galaxies with some residual effects of star formation. A totally independent measure of this is to use data from the {\it WISE} mid-infrared catalogues. {\it WISE} (W2$-$W3) colour is known to be an indicator of star formation through dust reprocessed emission \citep{Jarrett2011,Cluver2014}. In particular, \cite{Kettlety2018} demonstrated that a limit on (W2$-$W3) removed many galaxies otherwise thought to be passive systems. \cite{Fraser2016} used similar constraints to search for passive spirals. The local nature of our sample ensures that {\it WISE} can detect all the galaxies in W3 even if they are passive \citep[cf. the SINGS galaxy sample in][]{Cluver2017}.

Fig. \ref{NUVrs_wise23} shows the (rectified) NUV-optical versus {\it WISE} colour-colour diagram for our various samples.\footnote{Note that in the GAMA database, {\it WISE} magnitudes have been converted to the AB system, rather than the Vega system used in the cited {\it WISE} based papers.} It is clear that many of the optical red sequence galaxies which had already been removed as potential star formers from their (NUV$-u$) colours (pink points) {\em also} have the (W2$-$W3) colours of star forming galaxies, as delineated by the grey points. In other words, the previously removed objects are indeed consistent with being low-level star formers. 

In addition there is a tail of NUV red sequence galaxies (i.e. ones which were sufficiently red in (NUV$-u$) for retention) but which have the red (W2-W3) colours of star-forming galaxies. Removing these (the red points) leaves 122 red sequence objects which have both $y > 0.3$ {\em and} (W2-W3)$_{AB} < -0.3$, the edge of the bulk of blue cloud objects \citep[cf.][]{Jarrett2011}. We refer to these as NUV+{\it WISE} red sequence objects and these are shown in black in Fig. \ref{NUVrs_wise23}.

Referring back to the dark red (black bordered) histogram in Fig. 2, representing this final selection, it is evident that while a few of the  bluer (NUV$-r$) objects have been removed by the {\it WISE} cut, the range for our `double checked' passive sample remains the same as before, (NUV$-r$) from 5.2 to 6.5.\footnote{As we have the spectral measurements available, we have also checked that our final sample galaxies have 4000~\AA $\:$ break measurements of the amplitude expected for passive galaxies, i.e. D4000 $> 1.5$ (Balogh et al. 1999).}

\begin{figure}
\includegraphics[width=\linewidth]{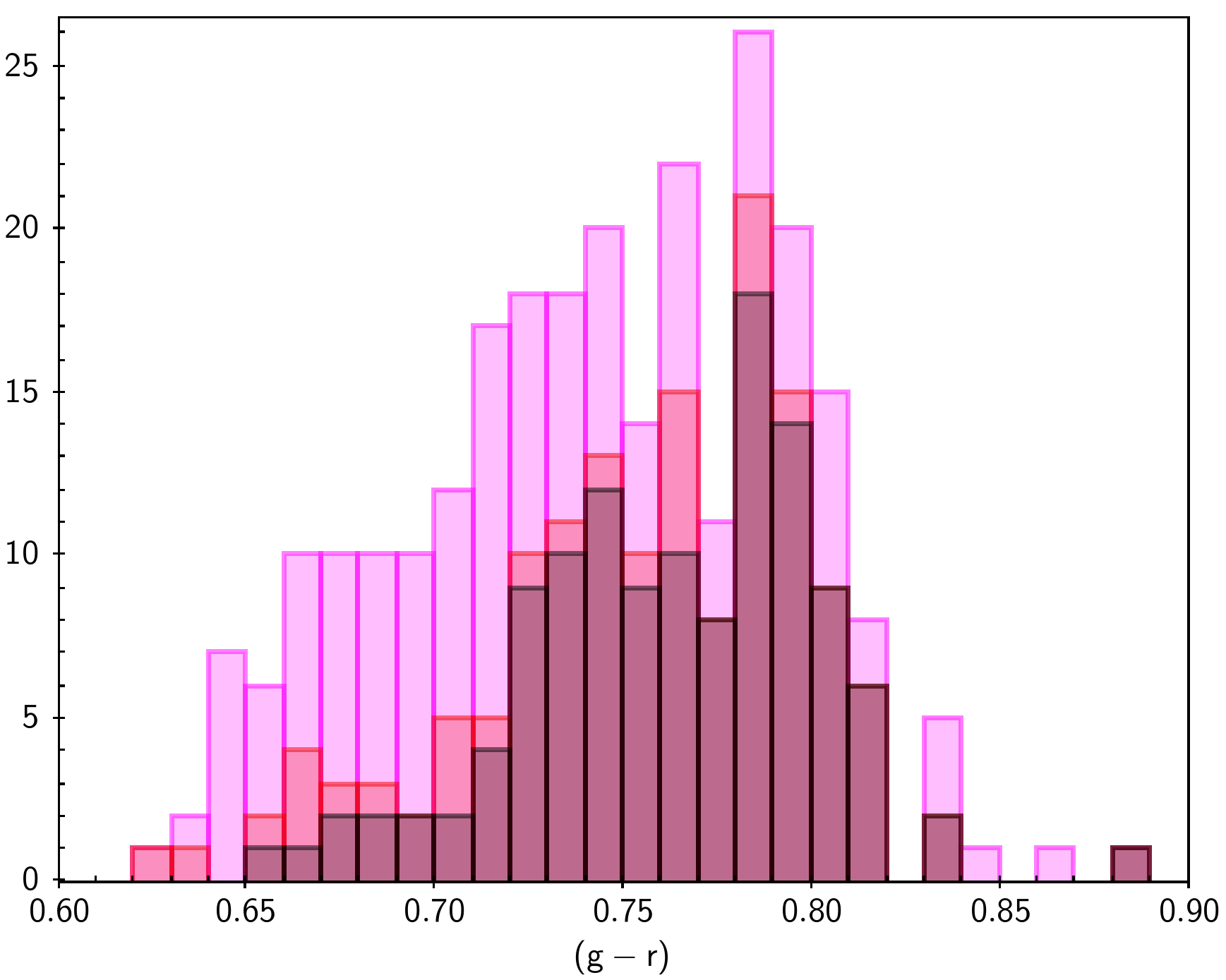}
\caption{Histograms of the optical rest-frame $(g-r)$ colours for the various red sequence selections, colour coded as before. The final NUV+WISE sample has a significantly narrower range compared to the original sequence, but simply choosing galaxies redder than, say, 0.72 would still include potential weak star formers (pink and red histograms) in addition to the members of our `best' sample (dark red histogram with black outlines). Photometric errors are around 0.03 magnitudes. }
\label{NUVWrs_gminusr_hist}
\end{figure}

In passing, it is interesting to note that if we return to the optical red sequence selection in $(g-r)$, the distribution (Fig. \ref{NUVWrs_gminusr_hist}) is much narrower for the galaxies which have survived our cuts than it was originally (now only about $\pm 0.05^{m}$). It might be supposed, therefore, that a more refined choice from the original $(g-r)$ versus $M_r$ colour magnitude diagram -- selecting only from the redder side of the evident red sequence -- might have achieved largely the same objective. However, the upper (redder) region of the optical red sequence does still include significant numbers of galaxies which are removed by our other cuts, mixed in with our best passive systems.  

As before, we can check the {\it WISE} mid-IR colours \citep{Cutri2013} of the Coma Cluster upturn sample from \cite{Ali2018a}. The reddest colours (so closest to the star-forming sequence) are at (W2-W3)$_{AB} \simeq -0.5$ \citep[transforming to AB magnitudes using values in][]{Jarrett2011}, again close to our cut value of $-0.3$. (We might, therefore, have made a slightly tighter cut, though this makes virtually no difference in practice, with only nine galaxies having (W2-W3)$_{AB}$ between $-0.5$ and $-0.3$).

\subsection{Specific Star Formation}

\begin{figure}
\includegraphics[width=\linewidth]{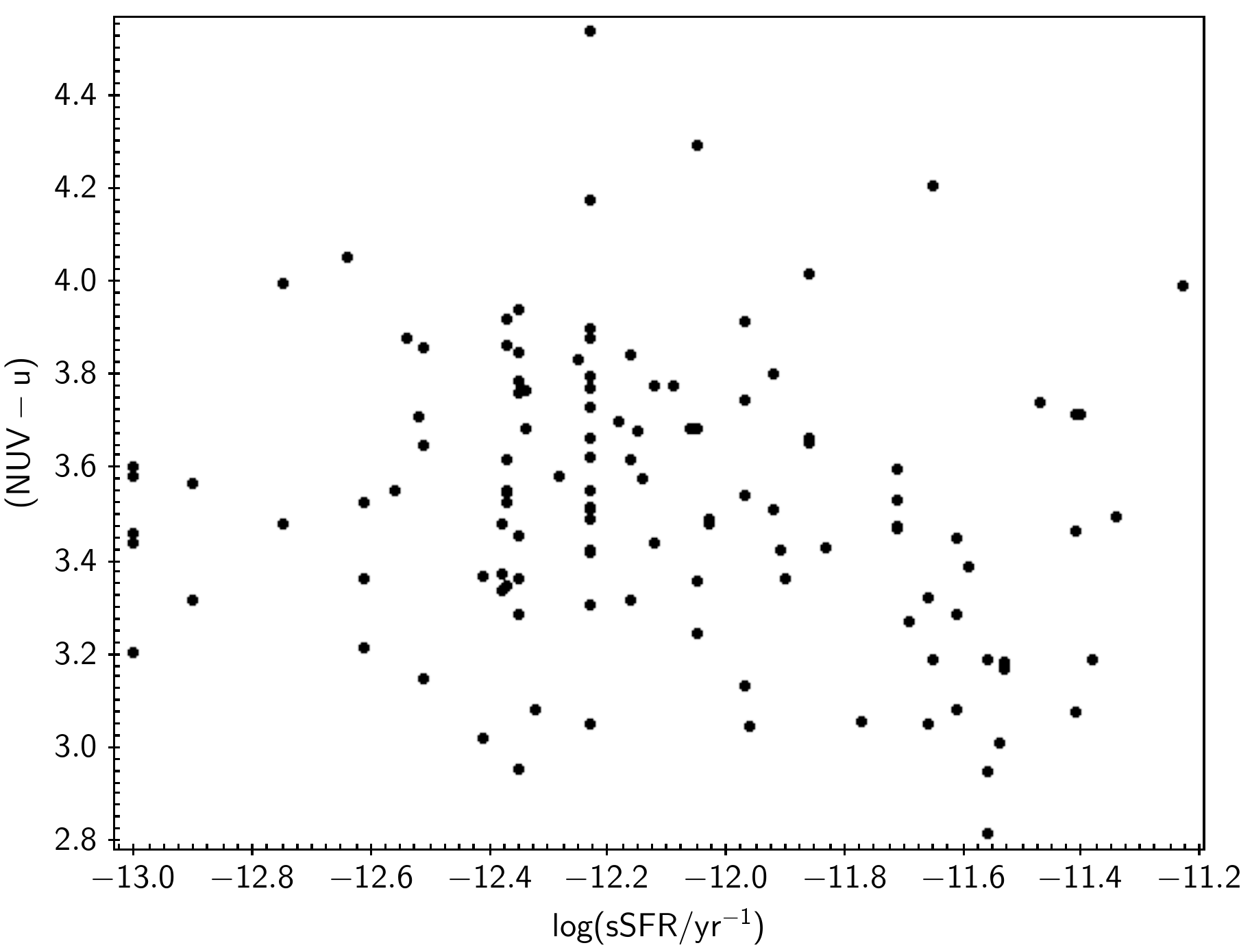}
\includegraphics[width=\linewidth]{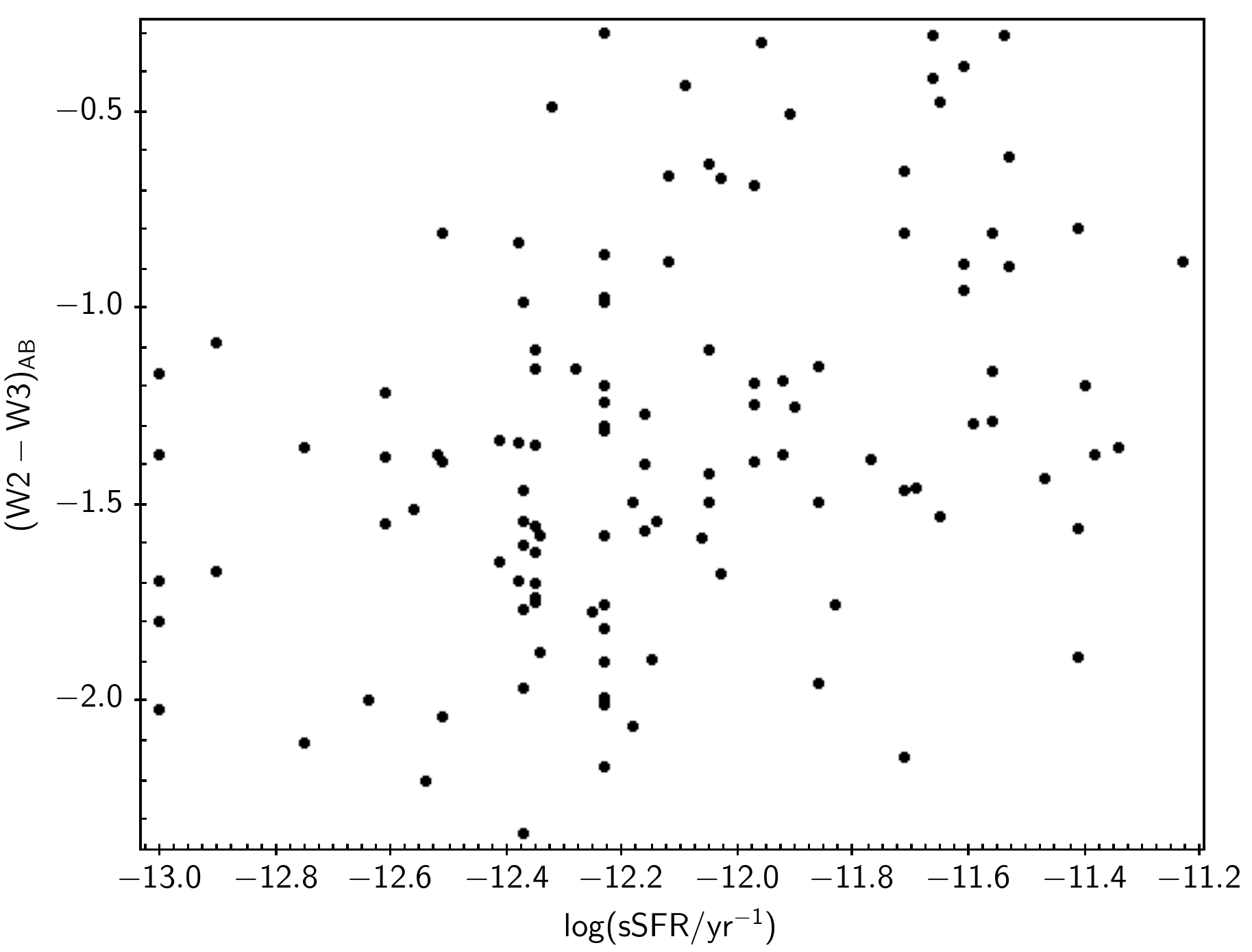}
\caption{MAGPHYS derived specific star formation rates (averaged over the last 0.1~Gyr) plotted against rest-frame (NUV-$u$) colour (top panel) and (W2-W3)$_{AB}$ (bottom panel) for our final NUV+{\it WISE} red sequence sample.}
\label{NUVWrs_nuv_ssfr}
\end{figure}

The MAGPHYS assigned sSFRs averaged over the last 0.1~Gyr \citep{Wright2016} for our final passive sample, while rather uncertain at such low levels\footnote{We can note, as validation of the MAGPHYS star formation rates, that they are consistent, even at low SFR, with those obtained from emission lines, where the latter are measurable (in the overall GAMA sample), see Wright et al. (2016) and Davies et al. (2016). For the star forming population, they also correlate as expected with the {\em WISE} colours.}, are virtually all below $10^{-11.4}$ yr$^{-1}$. Also the sSFRs do not correlate significantly with (NUV$-u$), and at best weakly with (W2$-$W3) (Pearson $r$ coefficients $-0.17$ and 0.32, respectively), in contrast to what would be expected if significant SFR were evident (see Fig. \ref{NUVWrs_nuv_ssfr}). Indeed, the large majority of the values are consistent with the lowest allowed value in the modelling, $10^{-13}$yr$^{-1}$, at the 3$\sigma$ level. Furthermore, as the models do not contain an upturn component, if one is present, the fitting will presumably attempt to assign any extra ultraviolet flux to star formation instead. In that case, the true sSFRs are likely to be even lower than quoted. 

Clearly, at these levels, the star formation rates cannot be expected to shift the (NUV$-r$) colours by a magnitude or more from those of the standard passive galaxy models (with no upturn) in order to cover the range of colours actually seen. To check this we can make a very simple calculation. We take the base (NUV$-r$) colour of a totally passive galaxy with no upturn to be 6.75, from the models of \cite{Conroy2009} \citep[see also][]{Ali2018b}, which effectively matches the upper limit for our data. Similarly for a star forming population, we take the bluest objects in our whole low $z$ sample, which have (NUV$-r) \simeq 1.5$ and sSFR $\simeq 10^{-9}$yr$^{-1}$. 
Scaling this component down and adding it to the old population such that the overall sSFR $\simeq 10^{-12}$yr$^{-1}$ (which is higher than that for the majority of our objects) we find a combined colour of 6.62, a change of only  $\simeq 0.1$ magnitudes. Thus we cannot produce the bluer (NUV$-r$) colours seen in our `passive' sample by adding the observationally permissable amount of star formation. In other words, even if {\em very} low-level star formation exists (which can not be ruled out), its effects are much smaller than those from the upturn populations. The general lack of emission lines in our sample galaxies similarly implies that AGN activity cannot be a significant factor in the UV fluxes.

\subsection{FUV}
We can also consider the FUV output from our galaxies, though this significantly reduces the number available; in particular only 58 of our 122 NUV+{\it WISE} red sequence galaxies are detected with FUV errors less than 0.2 magnitudes (i.e. 5$\sigma$ detections, as before).

\begin{figure}
\includegraphics[width=\linewidth]{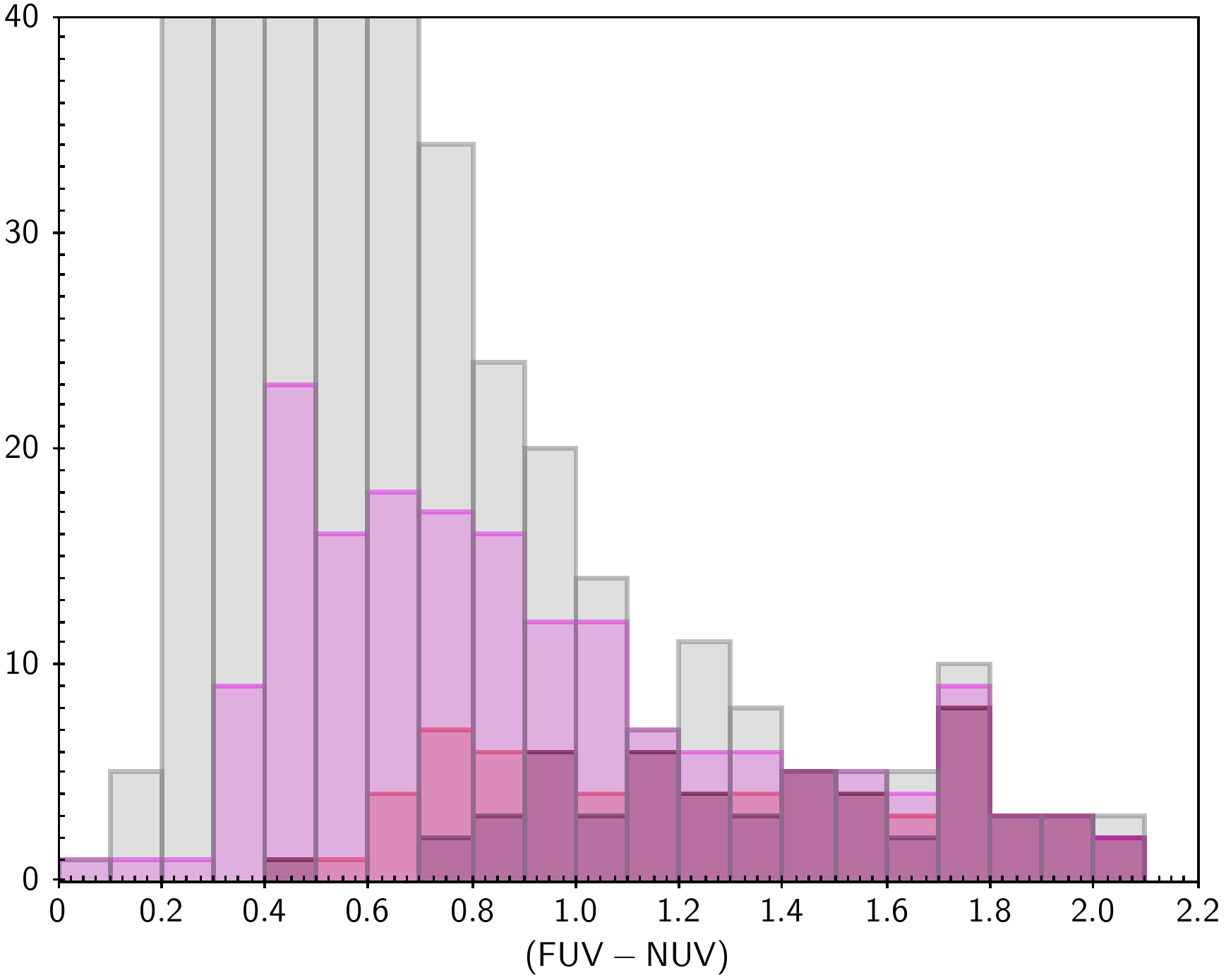}
\caption{Distributions of (FUV-NUV) colour (where available) for the various samples, plotted with the same colours as before. (Note that the vertical scale has been truncated to allow the details of the three red sequence selections to be made clearer). Star-forming galaxies (grey histogram) can be seen to be concentrated at (FUV-NUV) $<1.1$, as are the galaxies removed by our NUV and {\it WISE} colour cuts, while our NUV+{\it WISE} passive galaxies span the range $\sim 0.7$ to 2.1. Colour errors here are up to 0.3 magnitudes.}
\label{FUV_NUV}
\end{figure}

In Fig. \ref{FUV_NUV} we plot the distribution of (FUV-NUV) colours, where these are available, for galaxies in the various samples. It is evident that the star forming (blue cloud) galaxies (grey histogram) {\em and} most of the galaxies removed by our previous NUV and {\it WISE} cuts (pink and red histograms) are concentrated at (FUV-NUV) below 1.1  \citep[cf.][]{Brown2014}. The final NUV+{\it WISE} red sequence galaxies (dark red histogram) span (FUV-NUV) from 0.7 to 2.1 (plus some even redder outliers). Being extra cautious we could remove from our best sample, the objects with (FUV-NUV) colours overlapping those of the star formers (which is actually the reverse of the \cite{Yi2011} criterion, as they keep only the bluest galaxies as upturns). However, it turns out that this does not change the (NUV $-r$) distribution, as the NUV+{\it WISE} red sequence objects above and below, say (FUV-NUV) = 1.1, have the same range in (NUV$-r$). This supports the suggestion of \cite{Smith2012} that there is a continuum of colours for the passive upturn galaxies, spanning (fairly) blue to red in (FUV-NUV). It also agrees with the range in (FUV-NUV) colour found by \cite{Boselli2005} for (non-dwarf) early-type galaxies in the Virgo Cluster and by \cite{Brown2014} for their local spectroscopic sample. We therefore keep the NUV+{\it WISE} red sequence sample as our best passive sample in what follows, without any extra constraint from the FUV.

We can note, however, that for our passive galaxies there is an essentially linear correlation between (FUV$-r$) -- which undoubtedly measures the upturn population in passive galaxies with no star formation -- and the (NUV$-r$) colour which we use in the rest of this paper, albeit with a significant scatter of around $\pm 0.5^m$. Specifically, the bluest passive galaxies in (FUV$-r$), that is those with the strongest `classical' FUV-determined upturns, are also the bluest in (NUV$-r$). Conversely, the reddest galaxies in both cases match the colours expected for models of standard metal-rich old populations \citep[e.g.][]{Conroy2009} with no upturn. This correlation  between (FUV$-r$) and (NUV$-r$) is also seen for passive Coma Cluster galaxies with well characterised upturns in the sample of \cite{Ali2018a} and clearly supports our assertion that (NUV$-r$) is a good proxy for the upturn strength, even though also affected by the underlying old population \citep{Burstein1988,Dorman1995,Dorman2003,Smith2012,Ali2019}. \cite{Ali2018a} discussed in detail general two component SED models (i.e. with no star formation contribution), where the fits to the UV SED depend on both the metallicity and age of the old component and the strength (essentially the mass fraction of the relevant stellar sub-population) and temperature of the upturn component.


\section{Upturns and Environment}

Having established that (NUV$-r$) colour is a suitable proxy for upturn strength and defined our best sample of non-star forming galaxies, it is evident that these galaxies, our best candidates for purely passive objects, with or without an upturn, span the same range of (NUV$-r$) colours as found for passive galaxies in local clusters by \cite{Ali2018a}, who used more detailed ultraviolet SEDs which allowed them to rule out even very small amounts of star formation in their samples. \cite{Ali2019} further show that this colour range applies independently of the velocity dispersion or X-ray luminosity (essentially mass proxies) of the galaxies' host clusters (their Figure 4).

\begin{figure}
\includegraphics[width=\linewidth]{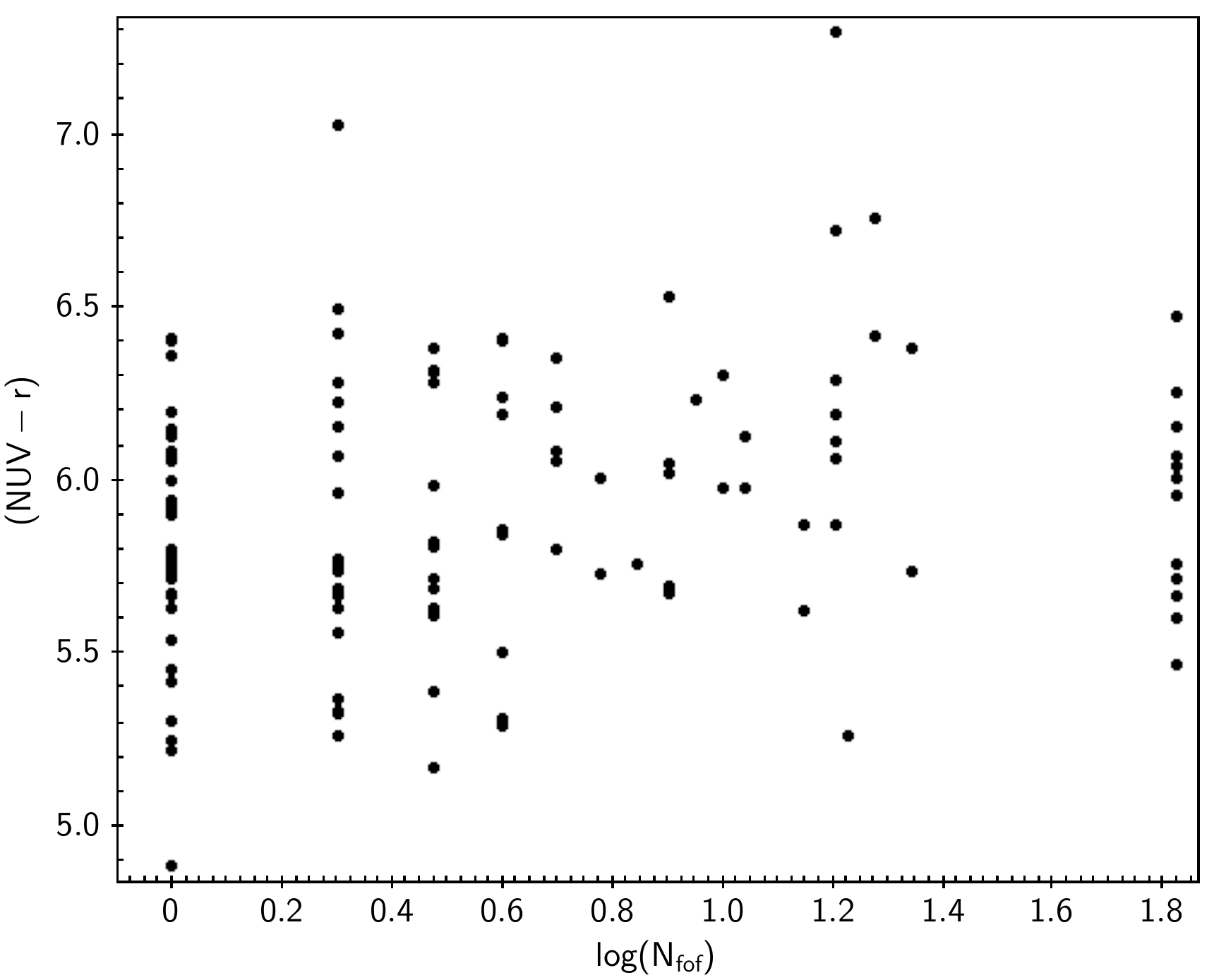}
\includegraphics[width=\linewidth]{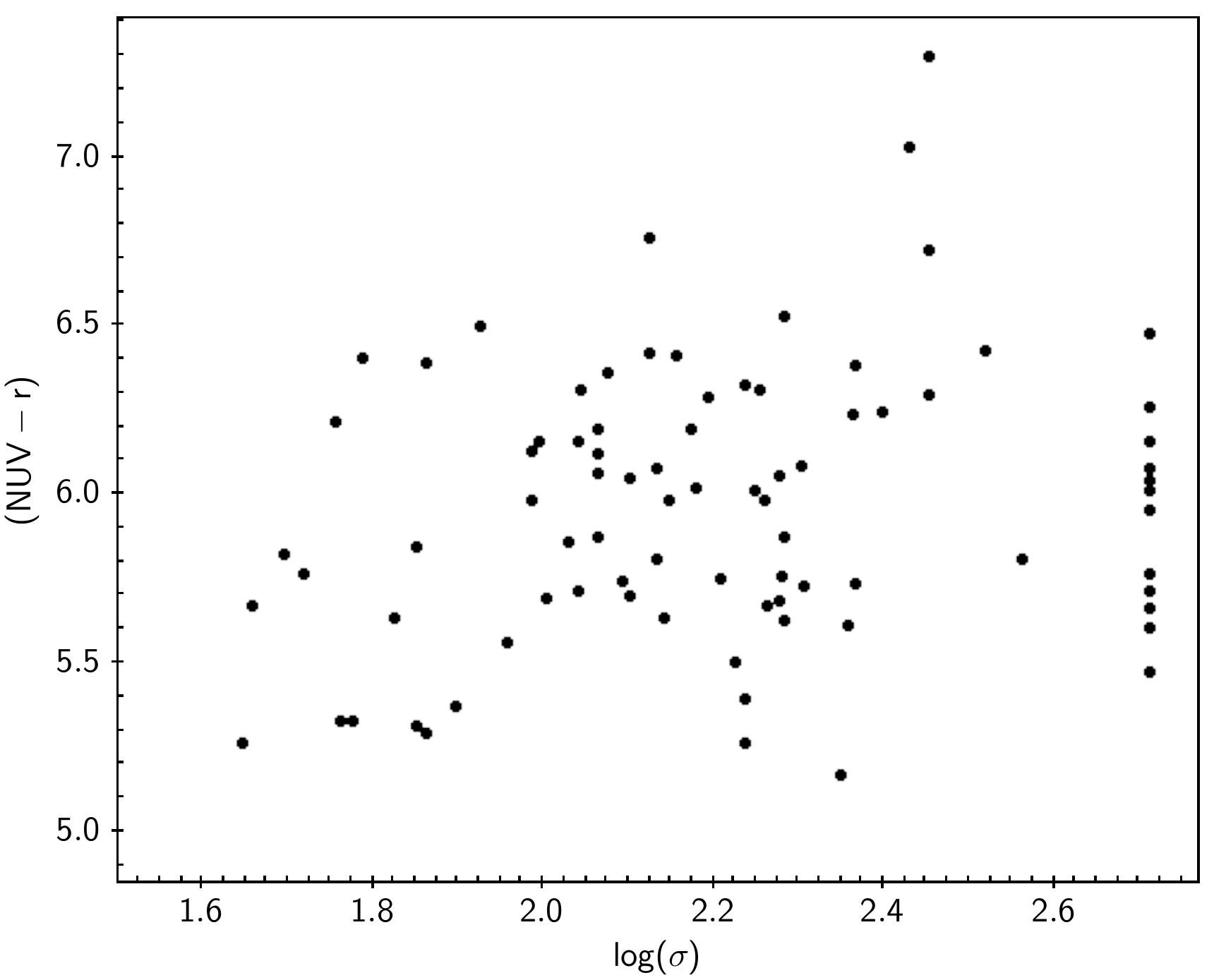}
\caption{Distribution of rest-frame (NUV$-r$) for the final passive (NUV+{\it WISE} red sequence) sample versus the multiplicity of the group containing the galaxy (top panel), and versus group velocity dispersion $\sigma$ (bottom panel). The bottom panel can be compared to the upper right panel of Figure 4 in Ali et al. (2019).}
\label{NUV_group}
\end{figure}

Fig. \ref{NUV_group} demonstrates the same independence of environment (Pearson r coefficient 0.15) among our small group galaxies (only one group has more than 25 members). The upper panel shows the (NUV$-r$) colours as a function of group (friends-of-friends) multiplicity from \cite{Robotham2011}. (Un-grouped galaxies are given a group multiplicity of 1, i.e. log(N$_{\rm fof}$) = 0). The lower panel matches the upper right panel of Figure 4 of \cite{Ali2019} and confirms that there is no dependence of the colour range on group velocity dispersion $\sigma$. This plot essentially continues the \cite{Ali2019} plot below their limit at $\sigma \sim 400$~kms$^{-1}$ with exactly the same colour range. Using either an estimated virial mass or a measure of the total stellar light from the group galaxies results in the same lack of dependence (r $<0.15$ in all cases). 

This strongly reinforces the conclusion of \cite{Ali2019} that the UV upturn is a phenomenon internal to individual galaxies and is not affected by the larger scale environment in which the galaxy finds itself. Note that this differs from the case for slightly bluer objects, with (NUV$-r) \leq 5$ and hence evidence for residual star formation, which {\em are} found to be environment dependent \citep[e.g.][see also Donas, Milliard \& Laget 1995]{Crossett2017}.

\begin{figure}
\includegraphics[width=\linewidth]{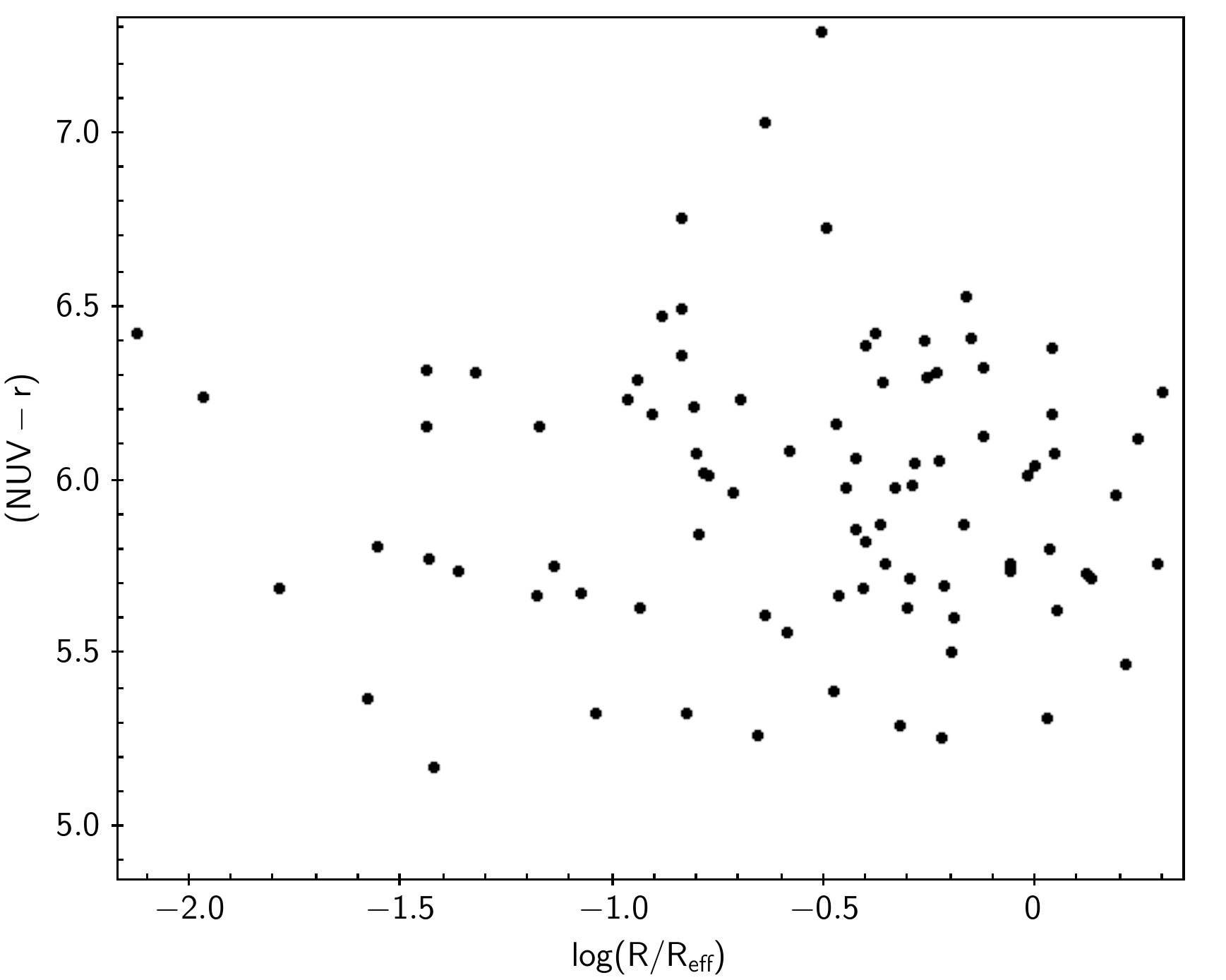}
\caption{Distribution of rest-frame (NUV$-r$) colour for the passive NUV+{\it WISE} red sequence sample as a function of radial distance $R$, normalised by host group effective radius, $R_{\rm eff}$. This can be directly compared to the upper right panel of Figure 5 in Ali et al. (2019).}
\label{NUVrs_r_reff}
\end{figure}

To complete the comparison with the \cite{Ali2019} analysis of the (NUV$-r$) colours, Fig. \ref{NUVrs_r_reff} shows the colour range as a function of radial distance from the group centre, normalised by the group's effective radius (the radius containing half the galaxies), again from \cite{Robotham2011}. Very small values are for galaxies near the geometric centre of their group: non-grouped galaxies are not plotted here, but we can already see their (matching) colour distribution from the previous plot (top panel, Fig. 8). 

As in the upper right panel of Figure 5 of \cite{Ali2019} for cluster galaxies, we see no positional dependence of the (NUV$-r$) colours among the group galaxies (r = 0.15). Galaxies close to and relatively far from the group centre are equally likely to show the broad range of (NUV$-r$) colours which evidences the UV upturn.

\begin{figure}
\includegraphics[width=\linewidth]{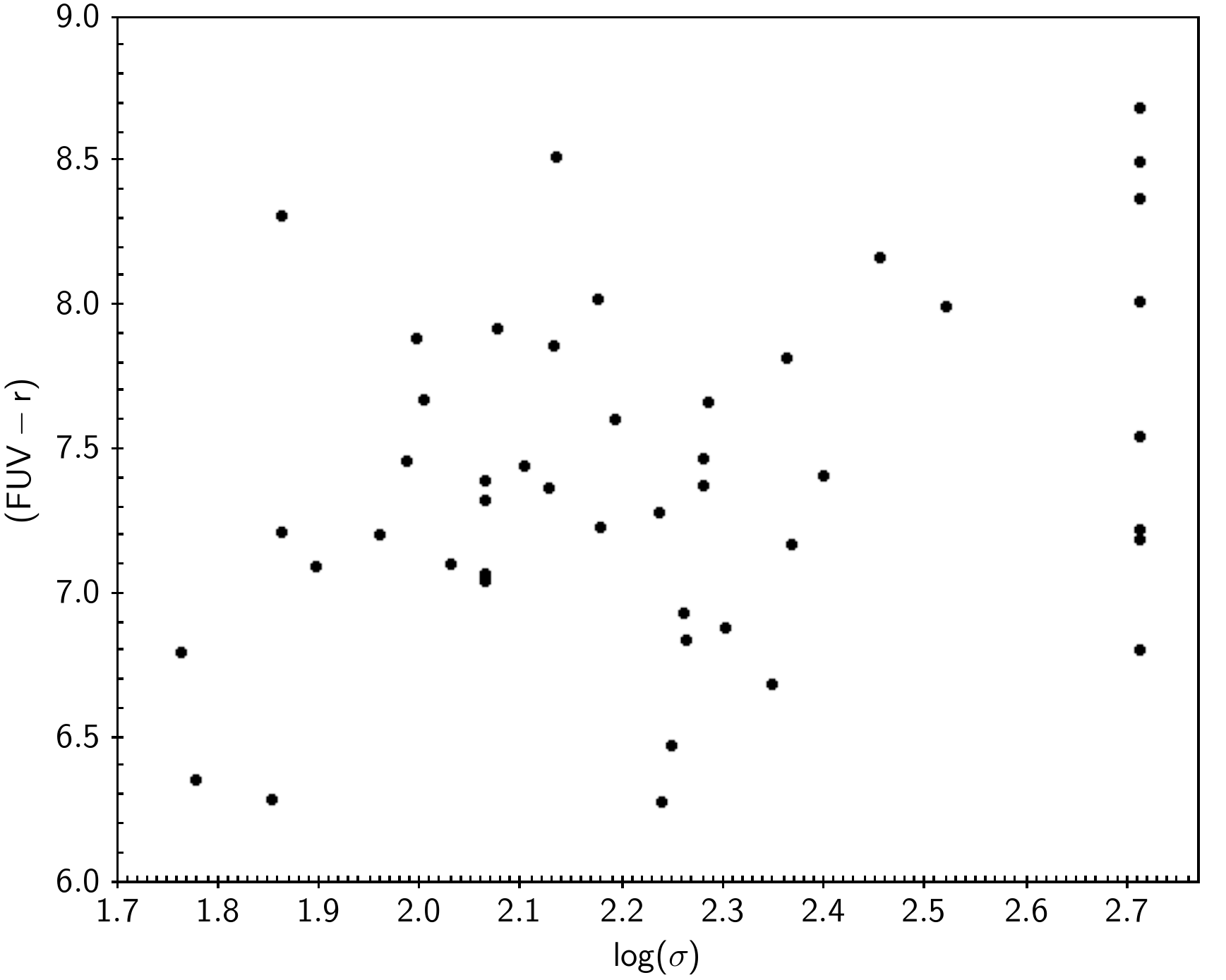}
\caption{(FUV$-r$) colour (where available) versus the host galaxy group velocity dispersion for the final NUV+{\it WISE} red sequence sample. This can be directly compared to the upper left panel of Figure 4 in Ali et al. (2019).}
\label{NUVWrs_FUV_sigma}
\end{figure}

\begin{figure}
\includegraphics[width=\linewidth]{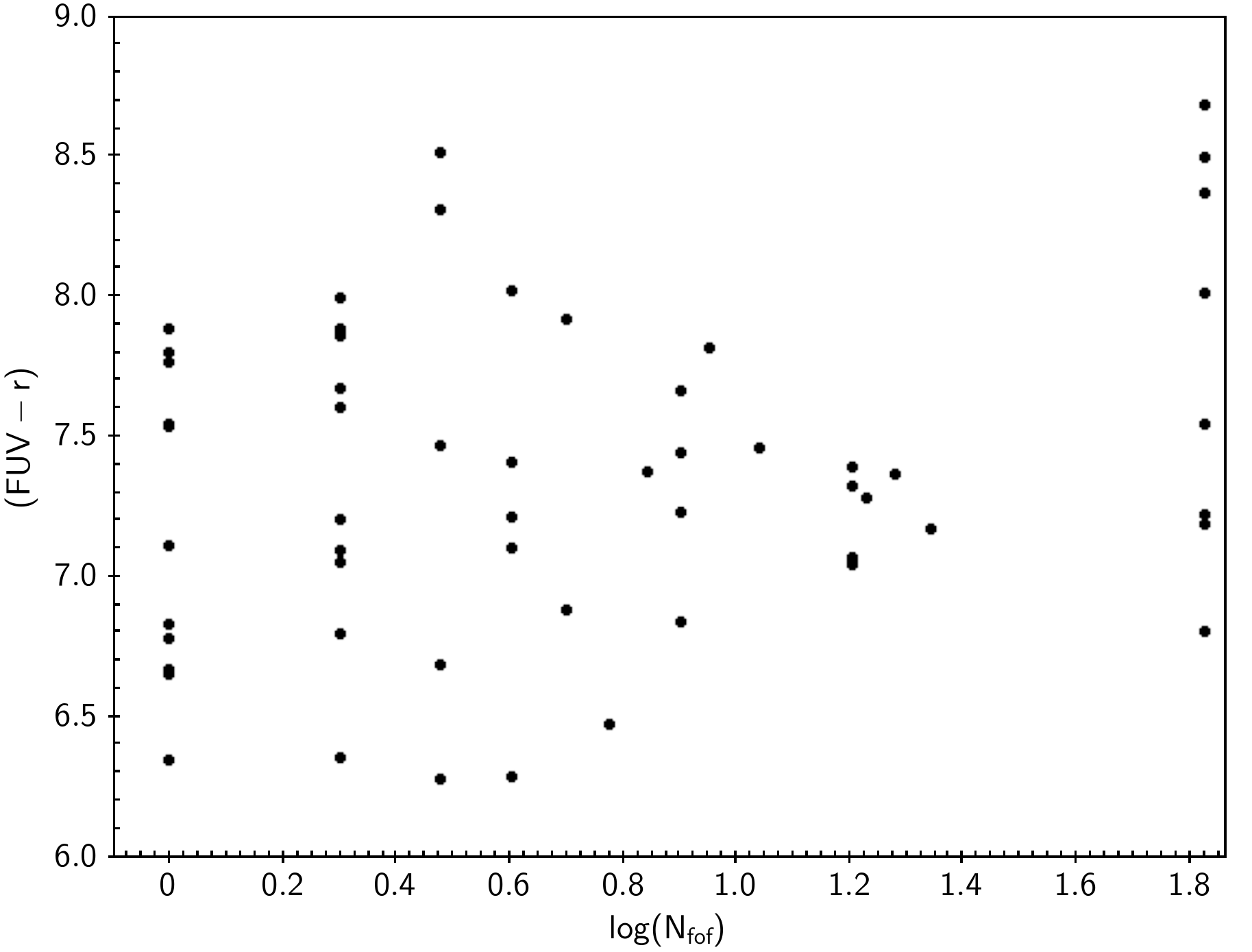}
\includegraphics[width=\linewidth]{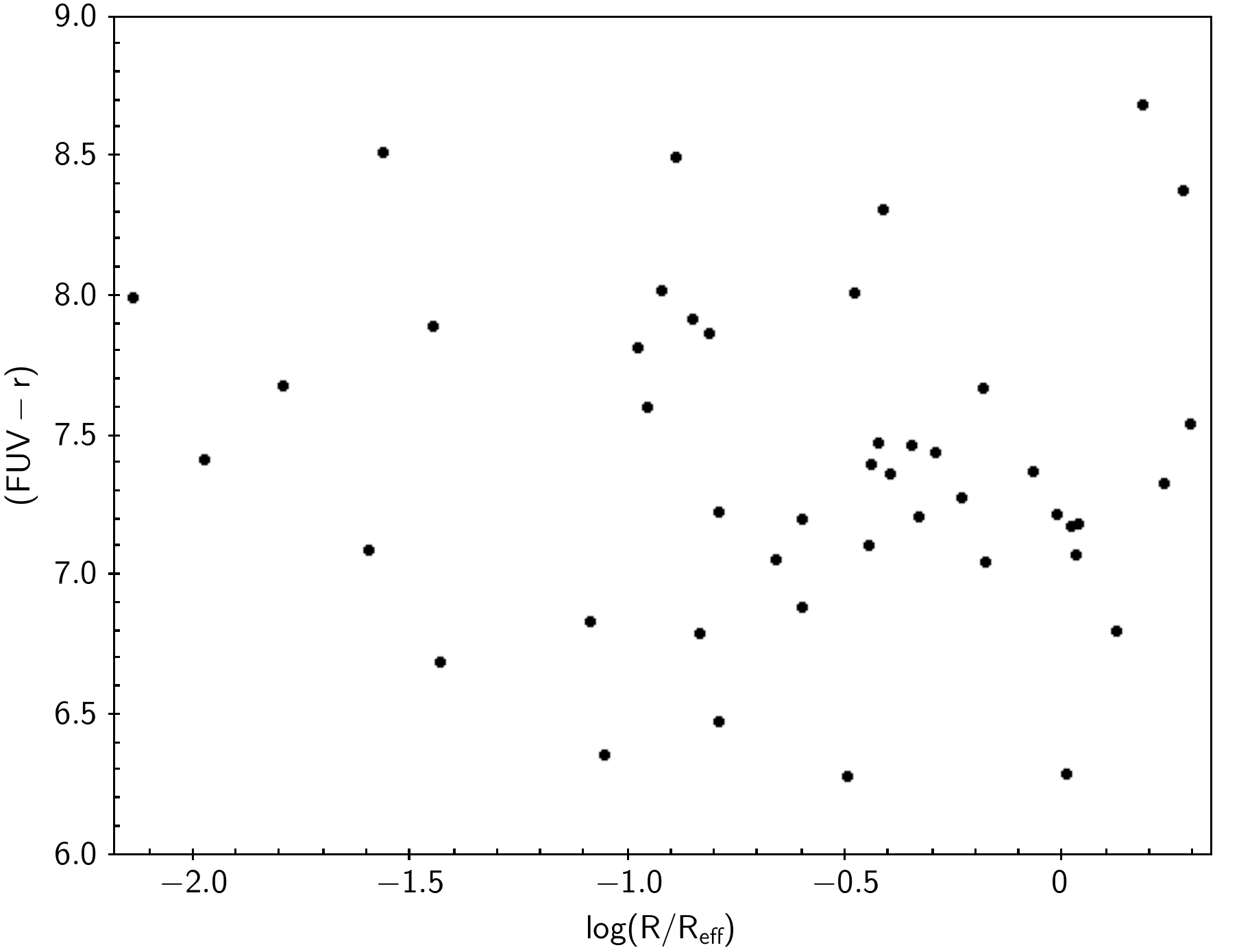}
\caption{(FUV$-r$) colour versus the host galaxy group multiplicity (top panel) and versus radial distance, normalised by group effective radius, (bottom panel) for the final NUV+{\it WISE} red sequence sample.}
\label{NUVWrs_FUV_env}
\end{figure}

We can repeat the exercise for the (FUV$-r$) colours of our sample galaxies, though the numbers with FUV detections are more limited. As an example, Fig. \ref{NUVWrs_FUV_sigma} shows the distribution of the FUV-optical colour versus the velocity dispersion of the galaxies' host groups. As with the NUV, we again see a wide spread of colours but no correlation (r = 0.04) of FUV-optical colour with environment \citep[see also][]{Atlee2009}. FUV-optical colour against group multiplicity and radial position, Fig. \ref{NUVWrs_FUV_env}, also show no effects ($|$r$|$ $< 0.1$), again in agreement with the results of \cite{Ali2019} for larger clusters. Thus the FUV data in Fig. \ref{NUVWrs_FUV_sigma} and Fig. \ref{NUVWrs_FUV_env} independently supports the lack of environmental dependence of the upturn, regardless of any question of the contribution to the NUV fluxes, as used in the rest of the paper, from non-upturn components.

\section{Additional Properties of Upturn Galaxies}

As we have a wide range of other properties measured for our GAMA galaxies, it is of interest to make use of these to explore whether our apparently passive galaxies with relatively blue UV-to-optical colours are distinguished in any other way.

If we make use of the optical morphologies from \cite{Moffett2016} we find that all our NUV+{\it WISE} red sequence sample galaxies are classified either as E or S0/a, in essentially equal numbers. The range of (NUV$-r$) is the same in each case. \cite{Bureau2011} tentatively suggested that UV upturns were preferentially in slow rotators, but a substantial number of S0 galaxies also containing upturn populations may argue against this. 

\begin{figure}
\includegraphics[width=\linewidth]{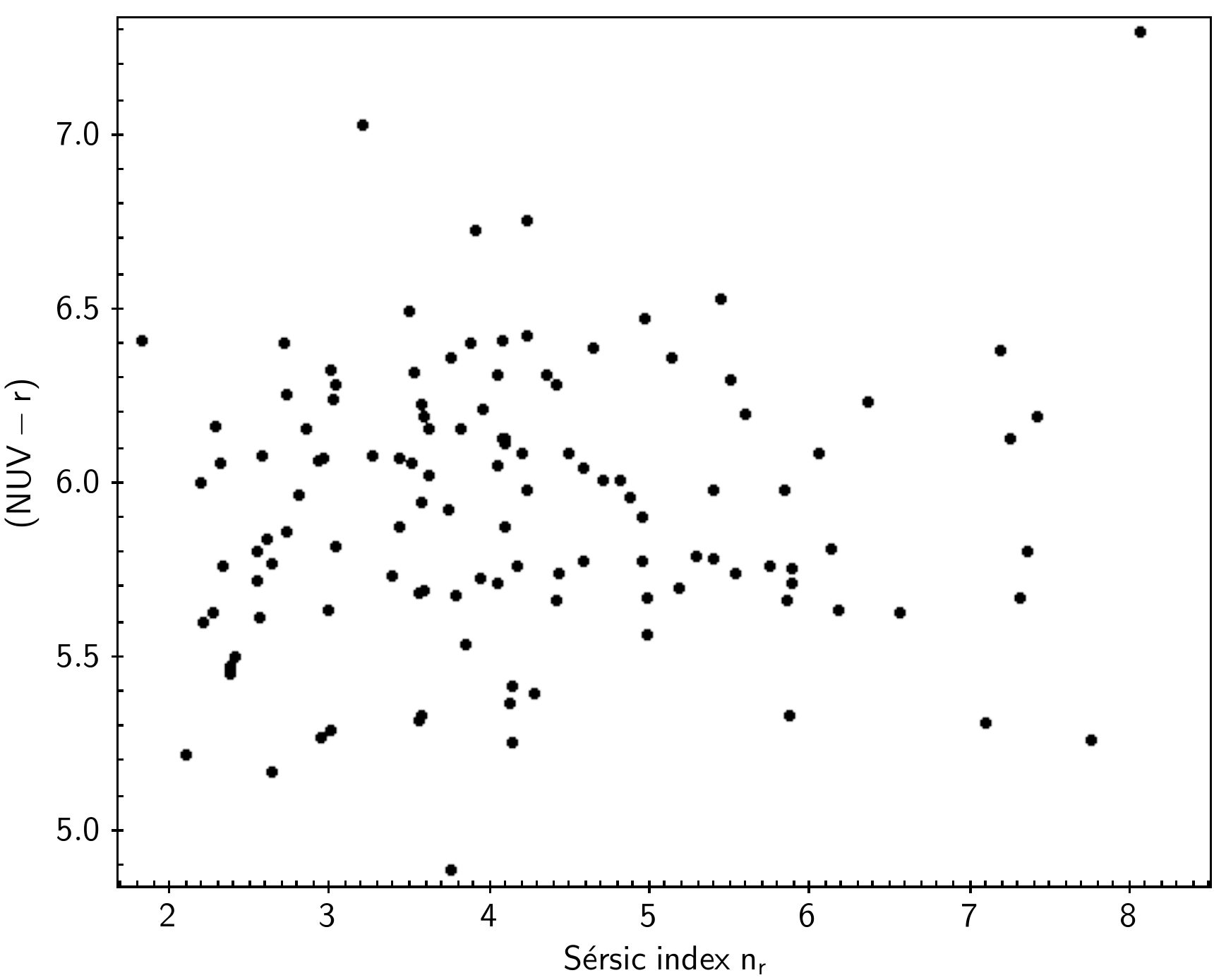}
\caption{Distribution of rest-frame (NUV$-r$) colour for the NUV+{\it WISE} red sequence galaxies as a function of $r-$band S\'{e}rsic index.}
\label{RS_sersic}
\end{figure}

Using the S\'{e}rsic surface brightness profile fits from \cite{Kelvin2012} we find no correlation of the (NUV $-r$) colour of our final sample galaxies with their S\'{e}rsic indices in any optical band (e.g. Fig. \ref{RS_sersic}), their central surface brightnesses or their axial ratios (Pearson $|$r$|$ below 0.06 in each case). Thus, the presence of a strong upturn component does not seem to depend in any obvious way on the overall structure of the galaxy. This may possibly disfavour the alternative models for the upturn based on producing hot stars via envelope loss in close binaries \citep{Han2007}, if the binary fraction is dependent on, e.g., stellar density \citep{Lucatello2015}.

Visual inspection of our sample galaxies does not show any sign that major mergers play a part in the production of blue NUV colours. Also, regarding lower levels of disturbance, from potential past or more minor interactions, no relationship is found between the residuals from the smooth S\'{e}rsic models (as measured by the $\chi^2$ goodness of fit) and the (NUV$-r$) colours and therefore upturn strength, implying that such interactions have played no part either. Given our previous results, this is, of course, as expected, since the interactions a galaxy will suffer are dependent on its overall environment. 

Although we have many derived stellar population parameters from the \cite{Taylor2011} fits, it may not be entirely appropriate to consider correlations with these, as the models used in \cite{Taylor2011} do not include an upturn component. Nevertheless, to the extent that the models only use data longwards of 3000 \AA, which should be reasonably unaffected by a separate UV upturn component, we can tentatively search for correlations between (NUV $-r$), measuring the strength of the upturn, and, e.g., the stellar population `age' (time since the peak in star formation), the timescale $\tau$ on which the star formation decayed \citep[cf.][]{Phillipps2019}, or stellar metallicity. 

\begin{figure}
\includegraphics[width=\linewidth]{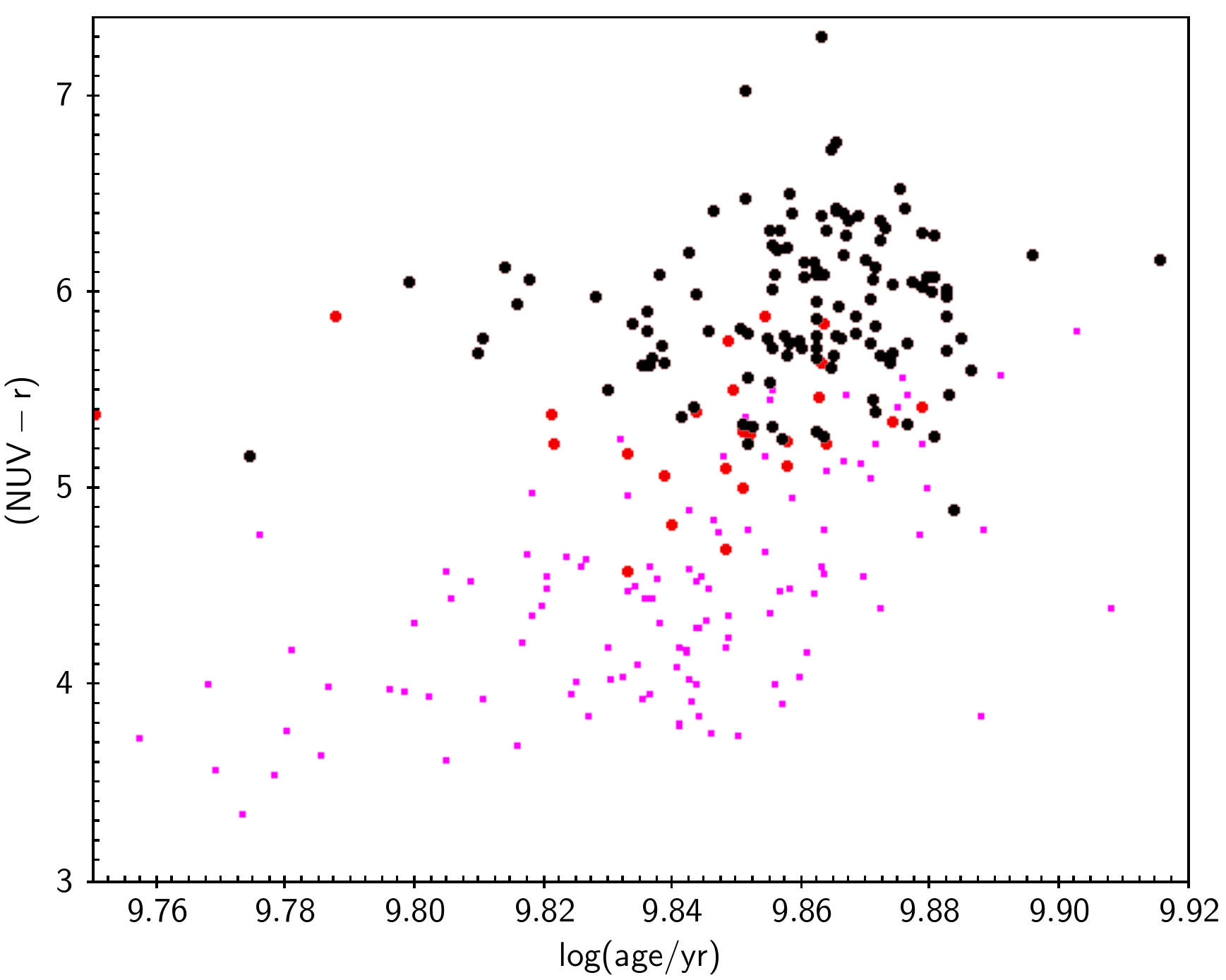}
\includegraphics[width=\linewidth]{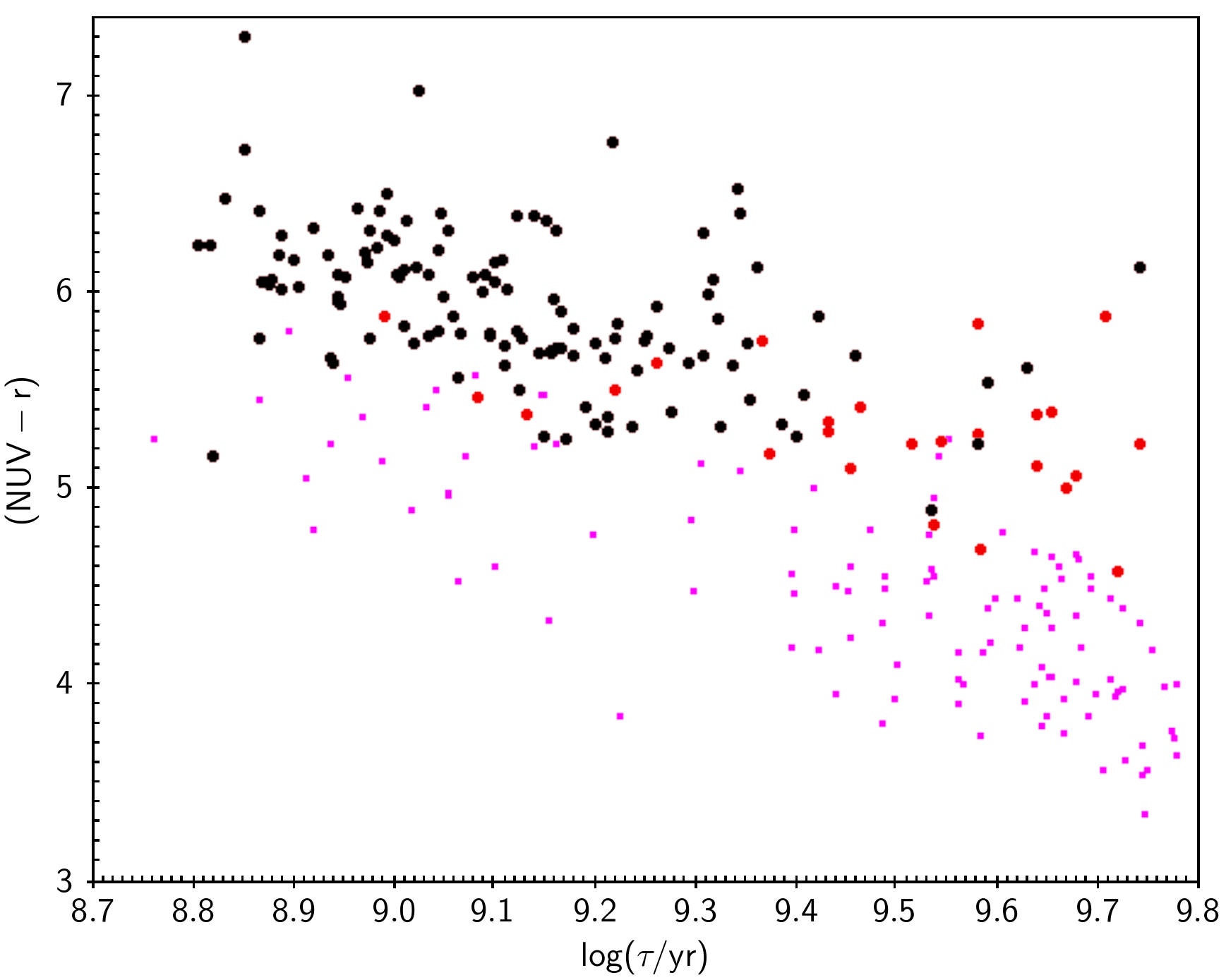}
\includegraphics[width=\linewidth]{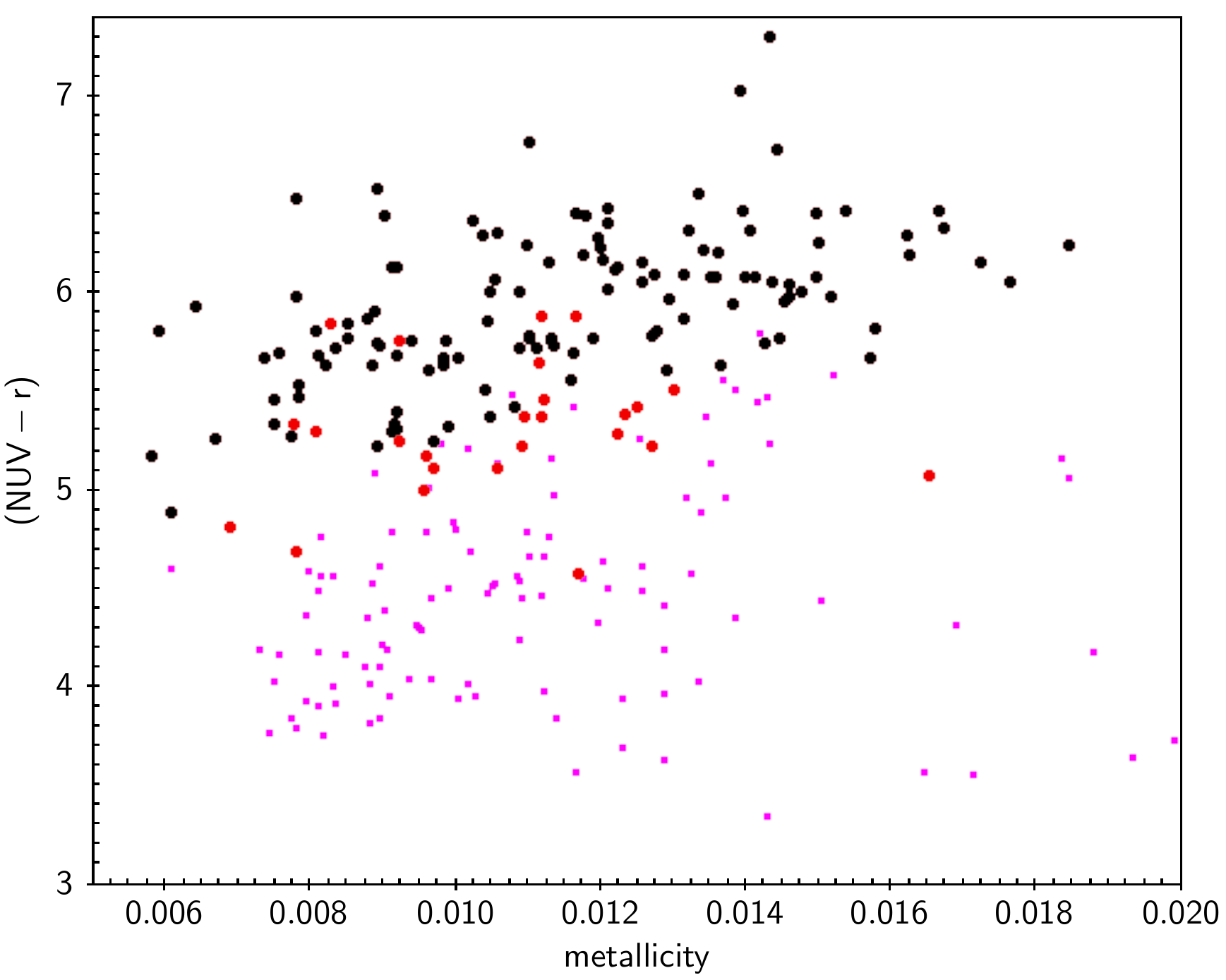}
\caption{Distribution of rest-frame (NUV$-r$) colour for the three red sequence subsamples (coloured as previously) as a function of (top) stellar population age, (middle) star formation timescale $\tau$, and (bottom) metallicity.}
\label{RS_properties}
\end{figure}

The top panel in Fig. \ref{RS_properties} shows that among our final sample (black points), there is no correlation between fitted age and the strength of the upturn (Pearson r = 0.09). However there is a clear trend for the optical red sequence galaxies which failed the (NUV $-u$) cut (pink points) to be modelled as having younger ages, as one might expect if they have (had) residual star formation. Note that these ages are determined by the total light, so do not necessarily reflect the age of the oldest stars, the UV upturn in particular being expected to result from HB stars at least 8 Gyr old \citep[e.g.][]{Kaviraj2007b,Ali2018c}.\footnote{A more precise measure of the oldest stellar population age might also be a useful discriminant between the He enriched models and alternative binary star evolution models, as the hot stars produced in the latter case need not be extremely old (Han et al. 2007, Li et al. 2013).}

For the star formation timescale, $\tau$ (middle panel), we again see that the bulk of the eliminated pink points (and most of the red circles, the galaxies which failed the {\em WISE} cut) tend to lie at larger values, consistent with longer lasting star formation, while the black points, our best (NUV+{\it WISE} red sequence) sample, generally ran down their star formation on timescales 1-2~Gyr. There is, here though, also a trend within the black points, with the tail of objects at (NUV$-r$) below 5.5 being among the galaxies with slightly longer $\tau$. As these bluer objects are also typically less massive, as discussed below, this may be a consequence of general `downsizing', i.e. smaller galaxies having more extended star forming lifetimes \citep{Cowie1996}. Unsurprisingly, given that the `age' $t$ is nearly constant for the passive galaxies, we see a corresponding distribution if we instead plot $t/\tau$ \citep{Phillipps2019}, again a measure of how evolved a galaxy is.

\begin{figure}
\includegraphics[width=\linewidth]{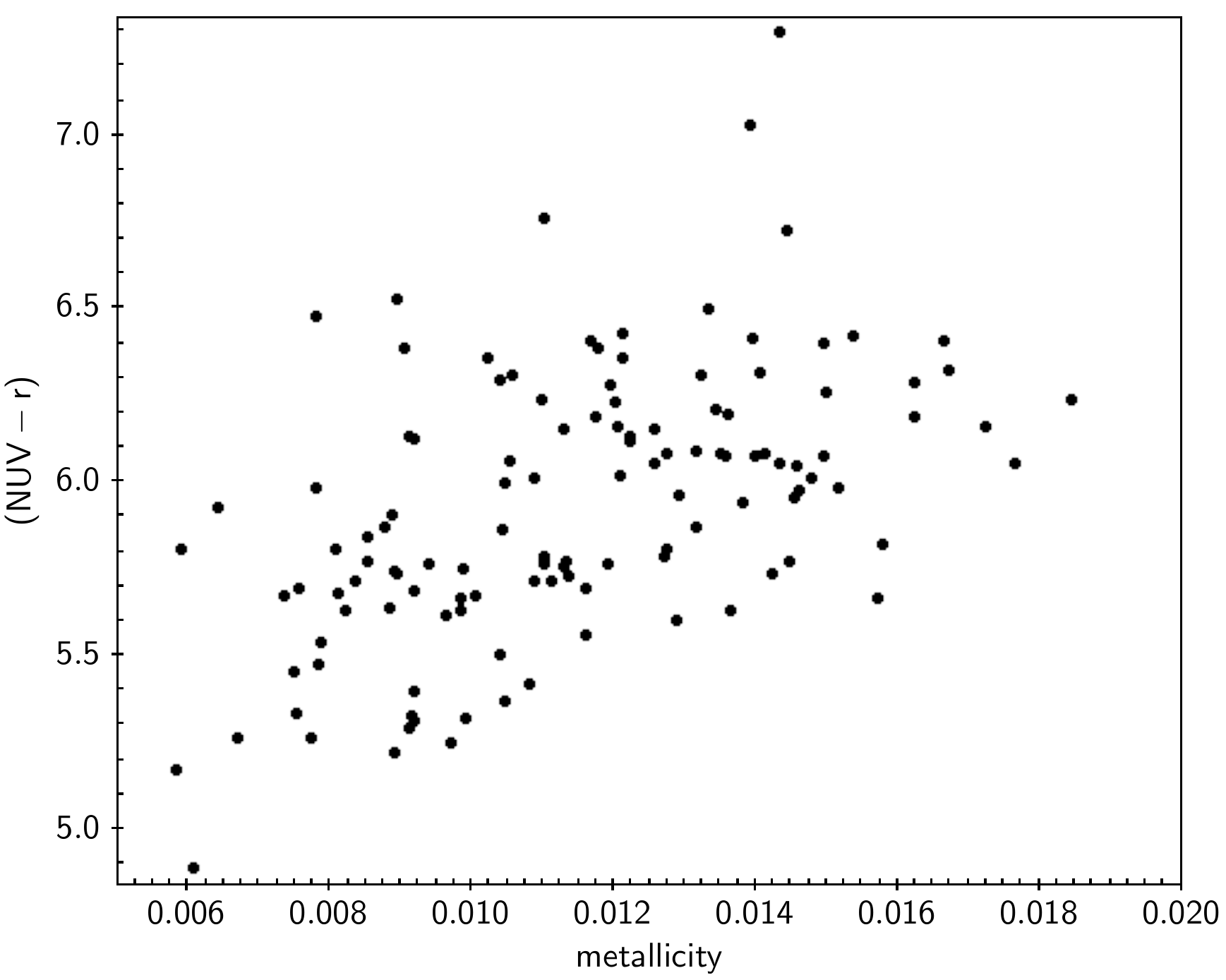}
\includegraphics[width=\linewidth]{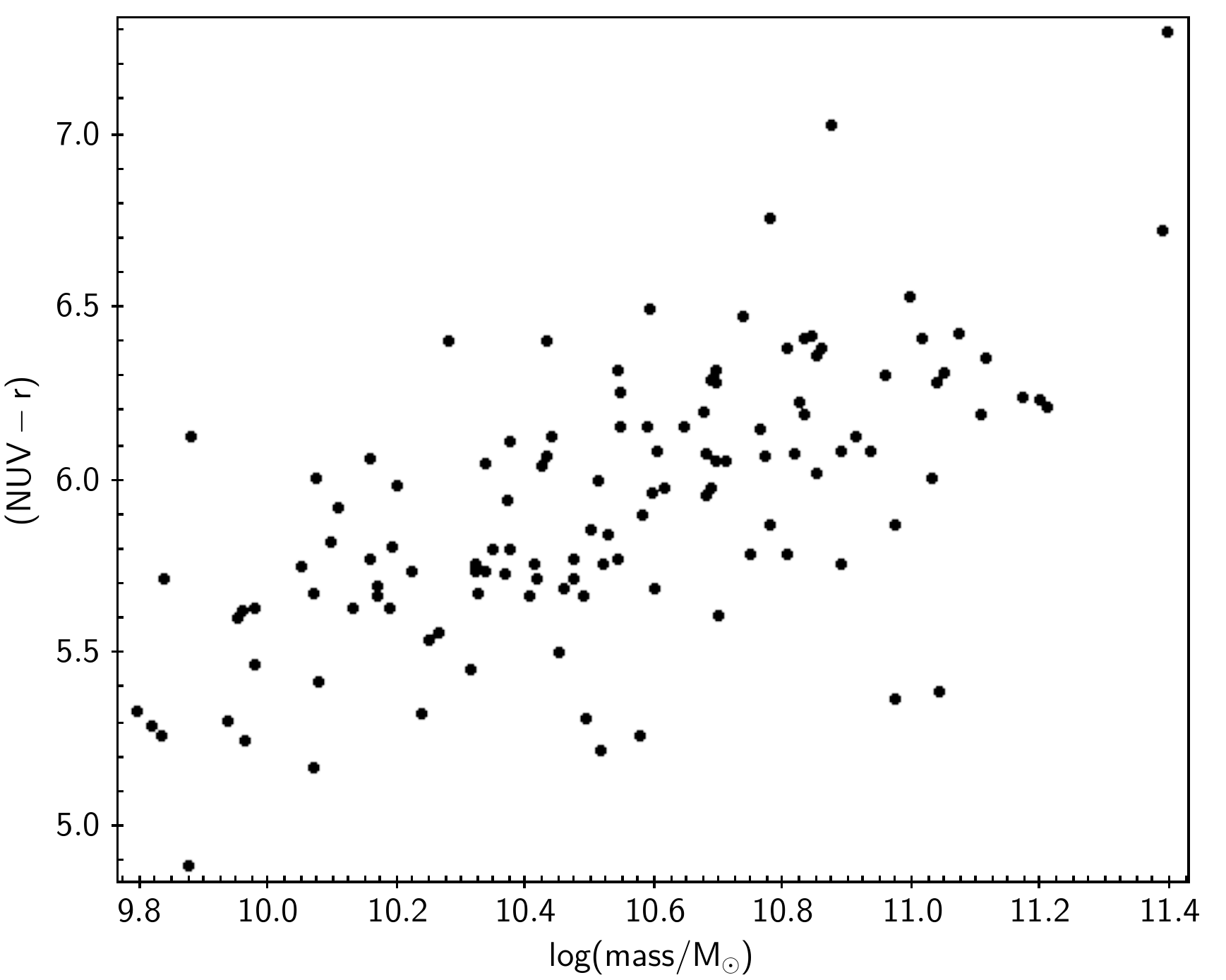}
\caption{Distribution of rest-frame (NUV$-r$) colour as a function of stellar metallicity as in the bottom panel of Figure 11, but for the final NUV+{\it WISE} red sequence only (top panel), and as a function of stellar mass for the same sample (bottom panel).}
\label{metal_mass}
\end{figure}

The stellar metallicities of our various red sequence sample galaxies are shown in the bottom panel of Fig. \ref{RS_properties}. Again, within the NUV+{\it WISE} red sequence objects (see the top panel of Fig. \ref{metal_mass} for an expanded version), while there is a clear overall correlation, the {\it width} of the (NUV$-r$) colour distribution is similar at all metallicities, suggesting a similar range of upturn components superimposed on base models whose colours vary with metallicity. The \cite{Conroy2009} models imply that (in the absence of an upturn component) at old ages passive galaxies with half-solar abundances should be about 0.6 magnitudes bluer in (NUV$-r$) than solar metallicity objects, which is consistent with the slopes of both the red and blue envelopes of our data. This agrees with the implications of the work of \cite{Dorman2003} who showed that the NUV SED will depend both on the strength of the upturn component and the metallicity of the old component (see their figure 1). The upper envelope in Fig. \ref{metal_mass} then matches the colours of purely old populations of the various metallicities, without an upturn component. Note that the wide range of (NUV$-r$) colours at each metallicity appears to argue against the suggestion by \cite{Schombert2016} that the upturn is {\em purely} a metallicity effect.

The few bluest (NUV$-r$) colours (below about 5.4) are all at the lower metallicity end and, unsurprisingly given the well known correlation of metallicity with mass \citep[e.g.][]{Faber1973,Gallazzi2006}, they are also at the low mass end of our sample (Fig. \ref{metal_mass}, bottom panel). From this plot, too, we see that there is a wide range of (NUV$-r$) at each mass \citep[see][for a similar plot in terms of luminosity]{Agius2013}.  

Given all the caveats -- especially that the population parameters are derived from a model which does not have an upturn component -- we therefore conclude that, within our best passive NUV+{\it WISE} red sequence sample, evidence for a correlation between the upturn stellar component and the underlying old stellar component is weak at best, {\em except} that lower mass (luminosity), lower metallicity passive galaxies extend to bluer UV-optical colours, and even this is likely due to the upturn component being superposed on a bluer background, rather than to a variation of the upturn component itself.

Finally, we make use of the MAGPHYS fits for the GAMA galaxies \citep{daCunha2008,Wright2016} for a number of ISM properties.\footnote{We should not use the MAGPHYS stellar population properties in this context, as they are derived from the full wavelength range including the FUV and NUV, while the template models do not include an upturn population. We can use them for the dust properties as these are determined primarily by the far-IR data from Herschel-ATLAS \citep{Eales2010}.} We find that among our final sample there is no correlation of blue upturn colour of our passive, early-type galaxies with any of dust mass, optical depth, temperature of the warm dust or temperature of the cold dust (Pearson $|$r$| < 0.15$ in each case). This is unsurprising given that the UV upturn is thought to be due to an old stellar population, so would not be expected to correlate with conditions in the present day ISM. We currently have no measurements of the neutral or molecular gas content of our sample galaxies.

\section{Discussion}

The relevance of our results to the question of UV upturns in early-type galaxies is two-fold. First, we have shown how to empirically `improve' a passive sample of galaxies, selected via the red sequence on a standard optical colour-magnitude diagram, by utilising additional constraints based on other broad band colours. Second, utilising our best passive galaxy sample derived in this way, we show that the range of (NUV$-r$) colours -- which we interpret as reflecting the strength of an upturn components -- is the same in all environments. 

As our first additional constraint we used the (NUV$-u$) colour, essentially the slope of the spectrum shortward of 3000 \AA. Specifically, from the distribution of galaxies in the (NUV$-u$) versus $(u-g)$ plane, we were able to extrapolate the locus of star forming galaxies and hence remove galaxies with residual star formation from the initial optical red sequence sample. By then using {\it WISE} (W2$-$W3) colours as an additional discriminant, we were able to further refine our non-star-forming sample. Within our final sample we find that galaxies with (NUV$-r) <5$ have been eliminated, verifying {\it post-hoc} the frequent assumption that this value can be used as a convenient limit for upturn galaxies versus galaxies with residual star formation. This process also removes some galaxies with (NUV$-r) >5$ which probably still have some star formation. Thus, as is well-known, unless more extensive data in the UV or IR is available, typically selected `optically passive' samples, such as those we require in order to search for upturn galaxies, will not be entirely pure and even a selection based on (NUV$-r$) may still leave some galaxies with residual star formation. Our final sample contains only around half (122) of the original (265) galaxies selected to be on the optical red sequence in rest-frame ($g-r$). This is consistent with the work of \cite{Mahajan2019} who find that around half of `red' galaxies in their GAMA sample have spiral, rather than early-type, morphologies \citep[see also][]{Kaviraj2007a,Crossett2014}.

Nevertheless, our `best' passive sample demonstrates that early-type galaxies {\em do} still have a wide range of (NUV$-r$) colours even when the most stringent limits are placed on their residual star formation.

Additionally, it may be that we have removed galaxies which do have upturn components but which are confused by ongoing or recent star formation (i.e. they contain two UV emitting components). Note, though, that the (NUV$-r$) distribution of our final sample is quite steep on the blue side, which may indicate that few upturn galaxies with no star formation are lost. In addition, we suggest that further constraints using the FUV may not be appropriate. 

Making use of our `best' passive sample, derived in this way, our major result is then that (within the relatively low density environments sampled by the GAMA data) we find the {\em same} range of (NUV$-r$) upturn colours, from 5.2 to 6.5, irrespective of the environment, as measured by group multiplicity, velocity dispersion, or halo mass, or by position within the group. We find the equivalent result using a more limited sample of galaxies with sufficiently accurate (FUV$-r$) colours. This complements and extends to lower halo masses (including isolated galaxies), the results of \cite{Ali2019}, who find no dependence of the range of upturn galaxy colours (and hence strength of upturn stellar component) with environment for galaxies in a wide range of nearby clusters.  Thus the stellar population responsible for the upturn appears to be entirely defined by internal stellar evolutionary processes within galaxies, not, for instance, by additional populations produced via mergers or other environment/cluster dependent effects. 

In addition, we find that the range of (NUV$-r$) colours is also independent (modulo the caveats noted above) of any (current) ISM properties, supporting the supposition that the upturn component is old. The overall galaxy (NUV$-r$) colours do depend on the stellar population metallicity, but the width of the distribution does not, implying the existence of UV upturn components with a similar range of strengths, superimposed on top of different underlying populations of varying (metallicity dependent) colour.  

\section*{Acknowledgements}

GAMA is a joint European-Australasian project based around a spectroscopic campaign using the Anglo-Australian Telescope. The GAMA input catalogue is based on data taken from the Sloan Digital Sky Survey and the UKIRT Infrared Deep Sky Survey. Complementary imaging of the GAMA regions is being obtained by a number of independent survey programs including {\it GALEX} MIS, VST KiDS, VISTA VIKING, {\it WISE}, {\it Herschel}-ATLAS, GMRT and ASKAP providing UV to radio coverage. 
 GAMA is funded by the STFC (UK), the ARC (Australia), the AAO, and the participating institutions. The GAMA website is http://www.gama-survey.org/. 

This work is based in part on observations made with the Galaxy Evolution Explorer (GALEX). GALEX is a NASA Small Explorer, whose mission was developed in cooperation with the Centre National d'Etudes Spatiales (CNES) of France and the Korean Ministry of Science and Technology. GALEX is operated for NASA by the California Institute of Technology, under NASA contract NAS5-98034.

This work made extensive use of TOPCAT \citep{Taylor2005} software packages, which are supported by an STFC grant to the University of Bristol.




\label{lastpage}

\end{document}